\documentclass[12pt,twoside,a4paper]{article} 
\usepackage[a4paper, total={6.5in, 8in}]{geometry}

\usepackage{amsmath}

\usepackage{draftwatermark}

\SetWatermarkText{DRAFT. Cite: Metodološki zvezki, Vol. 15, No. 1, 2018, 1–21. See: http://ibmi.mf.uni-lj.si/mz/2018/no-1/Cugmas2018.pdf.}
\SetWatermarkColor[gray]{0.5}
\SetWatermarkFontSize{0.5cm}
\SetWatermarkAngle{90}
\SetWatermarkHorCenter{20cm}

\usepackage[round]{natbib}
\usepackage{amsfonts}
\usepackage{multicol}
\usepackage{amsmath}
\usepackage[utf8]{inputenc}
\usepackage{caption}
\usepackage{dcolumn}
\usepackage{listings}
\usepackage{longtable}
\usepackage{color}
\usepackage{arydshln}
\usepackage{pbox}
\usepackage{algorithm}
\usepackage{algpseudocode}
\usepackage{blkarray, bigstrut}
\usepackage{subcaption}
\usepackage{makecell}
\usepackage[round]{natbib}
\usepackage{graphicx}
\usepackage[dvips]{epsfig}
\usepackage{index}
\usepackage{lhelp}
\usepackage{optparams}
\usepackage{psfrag}
\usepackage{enumitem}
\usepackage{hyperref} 
\usepackage{epstopdf}
\usepackage{pdfpages}
\usepackage{multirow}
\usepackage{hyperref}
\usepackage{rotating}
\usepackage{diagbox}
\usepackage{pdflscape}
\usepackage{pdfpages}
\usepackage{framed}

\usepackage{times}

\makeatletter
\@addtoreset{equation}{section}
\makeatother
\title{Comparing Two Partitions of Non-Equal Sets of Units} 

\author{
Marjan Cugmas\thanks{Faculty of Social Sciences, University of Ljubljana, Kardeljeva plo{\v{s}}{\v{c}}ad 5, 1000 Ljubljana, Slovenia; marjan.cugmas@fdv.uni-lj.si}
\and 
Anu{\v{s}}ka Ferligoj\thanks{Faculty of Social Sciences, University of Ljubljana, Kardeljeva plo{\v{s}}{\v{c}}ad 5, 1000 Ljubljana, Slovenia; anuska.ferligoj@fdv.uni-lj.si}
}

\date{ }
\begin{document}

\newtheorem{defn}{{\bf Definition}}
\newtheorem{thm}{{\bf Theorem}}
\newtheorem{cor}{{\bf Corollary}}
\newtheorem{pro}{{\bf Proposition}}
\newtheorem{lmm}{{\bf Lemma}}

\maketitle

\markboth{Cugmas and Ferligoj}
         {Comparing Two Partitions of Non-Equal Sets of Units}

\begin{abstract}
\date{ }
\maketitle
\setcounter{footnote}{1}
\cite{rand1971} proposed what has since become a well-known index for comparing two partitions obtained on the same set of units. The index takes a value on the interval between 0 and 1, where a higher value indicates more similar partitions. Sometimes, e.g. when the units are observed in two time periods, the splitting and merging of clusters should be considered differently, according to the operationalization of the stability of clusters. The Rand Index is symmetric in the sense that both the splitting and merging of clusters lower the value of the index. In such a non-symmetric case, one of the Wallace indexes \citep{wallace1983} can be used. Further, there are several cases when one wants to compare two partitions obtained on different sets of units, where the intersection of these sets of units is a non-empty set of units. In this instance, the new units and units which leave the clusters from the first partition can be considered as a factor lowering the value of the index. Therefore, a modified Rand index is presented. Because the splitting and merging of clusters have to be considered differently in some situations, an asymmetric modified Wallace Index is also proposed. For all presented indices, the correction for chance is described, which allows different values of a selected index to be compared.
\end{abstract}

\section{Introduction} \label{intro}

Many research problems require a comparison between partitions. One approach to doing this was proposed by \cite{rand1971} by introducing the Rand Index (RI). Since the expected value of the RI in the case of two random partitions does not take a constant value, \cite{hubertandarabie1985} suggested the Adjusted Rand Index (ARI). Even though there are many other similar indices \citep{albatineh2006}, the Adjusted Rand Index is one of the most often used indices for comparing two clusterings \citep{douglas2004}. Both the RI and ARI assume that two partitions are obtained from the same set of units.

Other well-known indices for comparing two partitions are the asymmetric Wallace Index B' and B'' (W1 and W2) \citep{wallace1983}, whose geometric mean is the Fowlkes and Mallows Index \citep{fowlkes1983method}. The main difference between the Wallace Index and the Rand Index is that merging and splitting of clusters have different effects on the value of the index. Here, merging of clusters is when the units from one or several clusters from the first clustering are clustered into one cluster in the second clustering. Simmilarly, we talk about splitting of clusters when the units from a given cluster from the first partition are clustered into two or several clusters in the second clustering. 

In this paper, we assume that two sets of units are not completely the same, but the intersection of these sets of units is a non-empty set of units. For example, when studying some social groups at two points in time certain participants might no longer be participating in the study at the second point (outgoers), some new participants might be recruited in the study in the second time period (newcomers) and other participants might be participating in both first and second time periods.

\begin{figure}[ht!]
  \setkeys{Gin}{width=1\textwidth}
\includegraphics[angle =0]{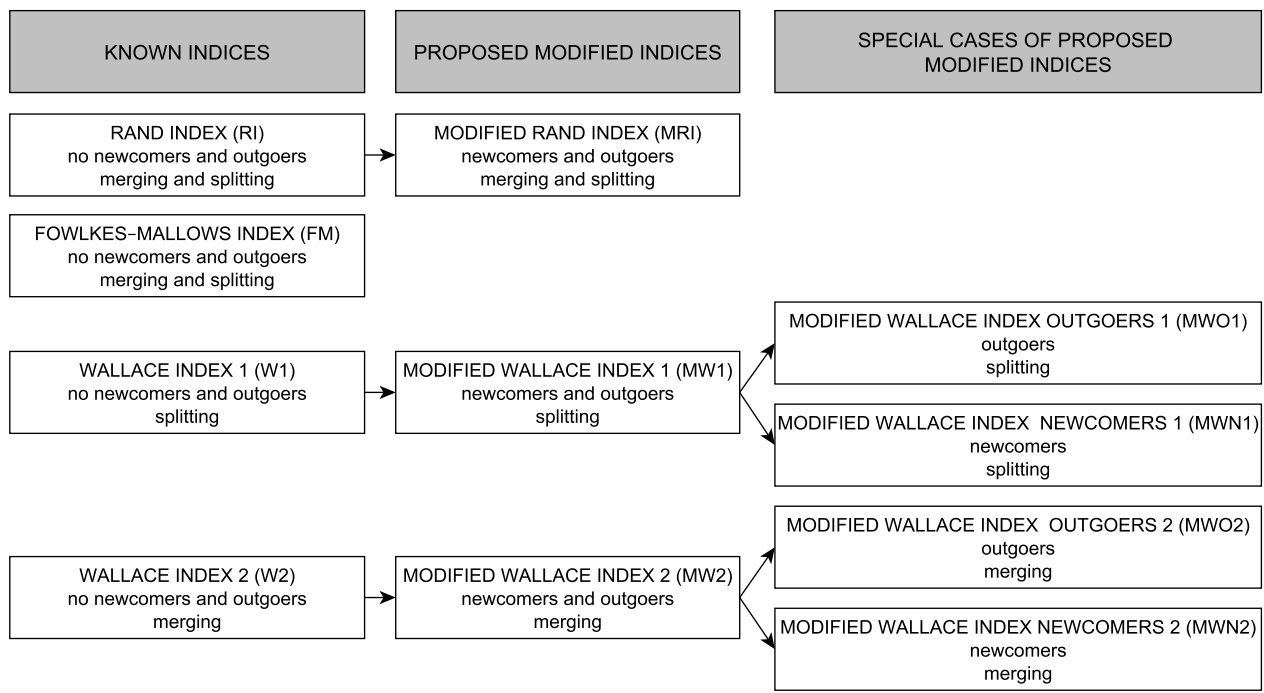}
  \caption{The list of well-known indices and the newly proposed indices (the general and the special cases) (the name along with the abbreviation is given and the assumptions about the presence of newcomers and outgoers is marked bellow along with the factors lowering the value of an index)}
  \label{scheme}
\end{figure}

Therefore, this paper proposes two indices for comparing two partitions obtained on two different sets of units. The first, the Modified Rand Index, is based on the original Rand Index and assumes that the stability of clusters is operationalized symmetrically, which means that both the splitting and merging of clusters have the same impact on the value of the index. The second index that is presented is based on the Wallace Index and assumes that the effects of splitting and merging clusters are operationalized differently: while the splitting of clusters indicates a lower value of the index, the merging of clusters does not. In the case of both indices, the number of outgoers and newcomers lower the value of the indices. If only newcomers or only outgoers are present (this means that one set of units for clustering is a subset of another set of units), the indices can be simplified. 

The choice of a given index depends on how the studied groups' stability is operationalized. For example, when studying the stability of research teams that publish scientific bibliographic units together at two periods of time (as in \cite{cugmas2016stability}), only the splitting of clusters and level of outgoers may be considered factors indicating a lower level of stability of the research teams since merging the research teams and newcomers leads to the formation of new scientific collaborations.

In the first part of the paper, all mentioned indices (see Fig.~\ref{scheme}) are presented in more detail. The second part discusses correction for chance, which is needed to compare two values of a selected index. One of the well-known non-parametric corrections for chance, which is based on simulations, is suggested for all proposed indices. The characteristics of all indices are illustrated through visualizations of misclassifications of units between two partitions. For this purpose, real data are used alongside generated data.

\section{The Rand Index, Fowlkes and Mallows Index and Wallace Index}

Given a set of units $S=\{O_1,...,O_n\}$, suppose $U=\{u_1,..., u_r\}$ and $V=\{v_1,..., v_q\}$ are two different partitions of $S$. The usual definition of a partition is that all clusters cover the whole set of units and that each pair of clusters does not overlap. Let $n_{ij}$ denote the number of units that are common to cluster $u_i$ and $v_j$. Then the cluster overlap between the two partitions $U$ and $V$ can be written in the form of a contingency table $\mathcal{M}$ where $n_{i\bullet}$ and $n_{\bullet j}$ are the number of units in clusters $u_i$ (row $i$) and $v_j$ (column $j$) respectively:

\begin{table}[H]
\caption{Contingency table $\mathcal{M}_{U  \times V}$}
\label{tab:ri_tab}
\centering
\begin{tabular}{clllllll}
\multicolumn{1}{l}{}           & \multicolumn{6}{c}{$V$}\\
\multirow{6}{*}{$U$}           & \multicolumn{1}{l|}{Class}     & $v_1$        & $v_2$        & ...                      & \multicolumn{1}{l|}{$v_{q}$}        & Sum \\ \cline{2-7} 
                               & \multicolumn{1}{l|}{$u_1$}     & $n_{11}$     & $n_{12}$     & ...                      & \multicolumn{1}{l|}{$n_{1q}$}       & $n_{1 \bullet}$     \\
                               & \multicolumn{1}{l|}{.}         & .            & .            &\multicolumn{1}{l}{.}     & \multicolumn{1}{l|}{.}              & .            \\
                               & \multicolumn{1}{l|}{.}         & .            & .            &\multicolumn{1}{l}{.}     & \multicolumn{1}{l|}{.}              & .            \\
                               & \multicolumn{1}{l|}{.}         & .            & .            &\multicolumn{1}{l}{.}    & \multicolumn{1}{l|}{.}              & .            \\
                                & \multicolumn{1}{l|}{$u_r$}     & $n_{r1}$     & $n_{r2}$     & ... & \multicolumn{1}{l|}{$n_{rq}$}   & $n_{r \bullet}$   \\ \cline{2-8} 
                                & \multicolumn{1}{l|}{Sum}       & $n_{\bullet 1}$             & $n_{\bullet 2}$           & ...                                & \multicolumn{1}{l|}{$n_{\bullet q}$}   & $n_{\bullet \bullet}=n$     
\end{tabular}
\end{table}

Based on the contingency table $\mathcal{M}_{U \times V}$, which is also called a matching table or cross-classification table in the field of cluster analysis, some very important quantities can be obtained. These are the basis for many indices used for comparing two partitions based on counting pairs (see \cite{albatineh2006} and \cite{Albatineh2011}, for some more examples) and are often presented in the form of a mismatch matrix: 

\begin{equation*}
  M=
  \begin{blockarray}{*{2}{c} l}
    \begin{block}{*{2}{>{$\footnotesize}c<{$}} l}
      pairs in same cluster in $V$ & pairs in different cluster in $V$ & \\
    \end{block}
    \begin{block}{[*{2}{c}]>{$\footnotesize}l<{$}}
      a & b & pairs in same cluster in $U$ \\
      c & d & pairs in different cluster in $U$ \\
    \end{block}
  \end{blockarray}
\end{equation*}

\noindent where the quantities $a$, $b$, $c$ and $d$ are sometimes denoted differently. These quantities are defined as follows:
\begin{itemize}
\item units in the pair that are placed in the same cluster in $U$ and in the same cluster in $V$:
\begin{equation} \label{eq:a}
a= \frac{1}{2} \sum_{i=1}^r \sum_{j=1}^q n_{ij}(n_{ij} - 1) = \frac{1}{2} {\Big (}\sum_{i=1}^r \sum_{j=1}^q n_{ij}^2 - \sum_{i=1}^r \sum_{j=1}^q n_{ij}{\Big )}
\end{equation}
\item units in the pair that are placed in the same cluster in $U$ and in different clusters in $V$:
\begin{equation} \label{eq:b}
b= \frac{1}{2} {\Big (}\sum_{i=1}^r n_{i \bullet}^2 - \sum_{i=1}^r \sum_{j=1}^q n_{ij}^2{\Big )}
\end{equation}
\item units in the pair that are placed in different clusters in $U$ and in the same cluster in $V$:
\begin{equation} \label{eq:c}
c= \frac{1}{2} {\Big (}\sum_{j=1}^q n_{\bullet j}^2 - \sum_{i=1}^r \sum_{j=1}^q n_{ij}^2{\Big )}
\end{equation}
\item units in the pair that are placed in different clusters in $U$ and in different clusters in $V$:
\begin{equation} \label{eq:d}
d= \frac{1}{2} {\Big (}n^2 + \sum_{i=1}^r \sum_{j=1}^q n_{ij}^2 - {\Big (}\sum_{i=1}^r n_{i \bullet}^2 + \sum_{j=1}^q n_{\bullet j}^2 {\Big )}{\Big )}
\end{equation}
\end{itemize}

\subsection{The Rand Index}

The Rand Index is often used to compare the simmilarity of two partitions and it is probably one of the most successful cluster validation indices. Its corrected-for-chance version called the Adjusted Rand Index (see Section~\ref{correction}) is not only used to evaluate two clustering methods, but can also be used, e.g. for link prediction in social network analysis \citep{Hoffman2015} or as a metric for evaluating supervised classification \citep{santos2009}.

It assumes one set of units for classification and two corresponding partitions. Based on the contingency table $\mathcal{M}_{U \times V}$, the Rand Index is defined as: 

\begin{equation} \label{eq:ri}
RI = \frac{a+d}{a+b+c+d} = \frac{\binom{n}{2} + \sum_{i=1}^r \sum_{j=1}^q n_{ij}^2 - \frac{1}{2} {\Big (} \sum_{i=1}^r n_{i \bullet}^2 + \sum_{j=1}^q n_{\bullet j}^2 {\Big )}}{{n \choose 2}}
\end{equation}

\noindent  and represents the proportion of all possible pairs that are in the same cluster and all possible pairs in different clusters in both partitions $U$ and $V$ compared to all possible pairs. 
When the two partitions are the same, the contingency table $\mathcal{M}_{U \times V}$ is a diagonal matrix. It is easy to show that the value of RI is 1 in such a case. The same is true when the size of each cluster is 1. If the number of clusters in partition $U$ equals the number of units $n$, then the value of the RI is approaching 1 when the number of clusters in partition $V$ is approaching the number of units $n$:

\begin{equation}
\lim_{q \to n} RI = \frac{{n \choose 2} + \frac{n}{2} - \frac{\sum_{j=1}^q n_{\bullet j}^2}{2}} {{n \choose 2}} = \frac{{n \choose 2} + \frac{n}{2} - \frac{n}{2}} {{n \choose 2}} = 1
\end{equation}

\noindent but when one partition consists of one cluster and another partition contains only clusters with a single unit (we say the two partitions are completely different), the value of RI is 0. For example, when the number of clusters in partition $U$ is 1 and the number of clusters in partition $V$ is approaching $n$, then the value of RI is approaching 0:

\begin{equation}
\lim_{q\to n} RI = \frac{{n \choose 2} - \frac{n^2}{2} + \frac{1}{2} \sum_{j=1}^q n_{1 j}^2} {{n \choose 2}} = \frac{{n \choose 2} - \frac{n^2}{2} + \frac{n}{2}} {{n \choose 2}} = 0
\end{equation}

The merging and splitting of clusters lower the value of $RI$. Let us assume the following partitions are obtained on the same set of units: $U$ with $r$ clusters, $V'$ with $q$ clusters, and $V''$ with $g$ clusters. Here, we assume that $q<g$ and clusters in partition $V''$ are obtained by the splitting of clusters from partition $V'$. In such case, the values $n_{ij}$ and $n_{\bullet j}$ in the contingency table $\mathcal{M}_{U \times V''}$ are lower than those in the contingency table $\mathcal{M}_{U \times V''}$. This results in lower values of $\sum_{i=1}^r \sum_{j=1}^q n_{ij}^2$ and $\sum_{i=1}^q n_{\bullet j}^2$ in the numerator in the definition of $RI$. Therefore, $RI(U, V') > RI(U, V'')$. Similar holds for the merging of clusters.

The difference between the Rand Index and the well-known simple matching coefficient\footnote{The list of some other selected association coefficients can be found in \cite{batagelj1995}.} \citep{sokal1958} is that the latter is calculated on the contingency table $\mathcal{M}_{2 \times 2}$, while the Rand Index is calculated based on the number of matching pairs presented in a mismatch table (see Eq. (\ref{eq:a}), (\ref{eq:b}), (\ref{eq:c}) and (\ref{eq:d})). \cite{warrens2008} has shown it is possible to calculate the Adjusted Rand Index by first forming the mismatch table and then computing Cohen's $\kappa$ index based on this table.

\subsection{The Fowlkes and Mallows Index}

\cite{fowlkes1983method} proposed the Fowlkes and Mallows Index ($FM$) which, like the Rand Index, is a measure of the similarity of two partitions obtained on the same set of units \citep{wallace1983} defined as: 

\begin{equation}
FM=\sqrt{\frac{a}{a+b} \times \frac{a}{a+c}}
\end{equation}

The main difference between the Fowlkes and Mallows Index versus the Rand Index is that the value of the Fowlkes and Mallows Index in the case of two random and independent partitions approaches 0 as the number of clusters (in partition $U$ or $V$) increases. As in the case of the Rand Index, the value of the Fowlkes and Mallows Index is still relatively high in the case of a small number of clusters \citep{wallace1983}. 

\subsection{Wallace indices}

\cite{wallace1983} pointed out that the symmetric measure FM is the simple geometric mean of two asymmetric indices:

\begin{equation}
W1 = \frac{a}{a+b}
\end{equation}

\begin{equation}
W2 = \frac{a}{a+c}
\end{equation}

Because $W1$ and $W2$ are asymmetric indices, the order of the partitions has to be considered (e.g. if two sets of units are obtained at two different time points, the partition corresponding to the first time point is denoted as $U$ and the partition corresponding to the second time point is denoted as $V$). $W1$ can be interpreted as the probability that, for a given data set, two units are classified in the same cluster in partition $V$ if they have been classified in the same cluster in partition $U$ (the splitting of clusters lowers the value of the index, while the merging of clusters does not lower the value) \citep{severiano2011adjusted}. Similarly, $W2$ is defined as the probability that, for a given data set, two units are classified into the same cluster in partition $V$ if they have been classified into different clusters in partition $U$ (the merging of clusters lowers the value of the index, while the splitting of clusters does not lower the value). 

\section{Modified indices considering newcomers and outgoers}

In this section, the newly proposed indices based on the Rand Index, the first Wallace Index ($W1$) and the second Wallace Index ($W2$) for measuring the similarity of two partitions are proposed. In comparison with the standard indices presented in the previous section, they do not only assume one set of units for classification. There can be two, where the intersection between two sets of units is a non-empty set of units. This means that, e.g. if the sets of units are observed at two different time points, there are some units which are present in both sets of units (the units in the intersection), some units which are in the first set of units but not in the second set of units (outgoers) and some units which are not in the first set of units but appear in the second set of units (newcomers).

Given two sets of units $S=\{O_1,...,O_s\}$ and $T=\{O_1,...,O_t\}$, where $S \cap T \neq \emptyset$, suppose $U=\{u_1,...,u_r\}$ is the partition of $S$ with $r$ clusters and $V=\{v_1,...,v_q\}$ is the partition of $T$ with $q$ clusters.  The newcomers $T\backslash S$ define a new cluster which is added to the partition $U$ ($U' = U \cup \{u_{r+1}\}$) and the out-going $S \backslash T$ define a new cluster which is added to the partition $V$ ($V' = V \cup \{v_{q+1}\}$). The number of units (denoted by $n$) is then equal in both partitions. The partitions can be presented in a contingency table $\mathcal{M}_{U' \times V'}$. If the cluster representing the newcomers is arranged in the last row, and the cluster of out-going units is arranged in the last column, we obtain $\mathcal{M}_{U' \times V'}$ contingency table (see Table~2).

\begin{table}[H]
\caption{Contingency table $\mathcal{M}_{U' \times V'}$}
\label{tab:rim_tri}
\centering
\begin{tabular}{clllll:ll}
\multicolumn{1}{l}{}           & \multicolumn{7}{c}{$V'$}                                                                                                                             \\
\multirow{7}{*}{$U'$}          & \multicolumn{1}{l|}{Class}     & $v_1$        & $v_2$        & ... & $v_{q}$          & \multicolumn{1}{l|}{$v_{(q+1)}$}    & Sum         \\ \cline{2-8} 
                               & \multicolumn{1}{l|}{$u_1$}     & $n_{11}$     & $n_{12}$     & ... & $n_{1q}$         & \multicolumn{1}{l|}{$n_{1(q+1)}$}   & $n_{1\bullet}$     \\
                               & \multicolumn{1}{l|}{$u_2$}     & $n_{21}$     & $n_{22}$     & ... & $n_{2q}$         & \multicolumn{1}{l|}{$n_{2(q+1)}$}   & $n_{2\bullet}$     \\
                               & \multicolumn{1}{l|}{.}         & .            & .            &     & .                & \multicolumn{1}{l|}{.}          & .            \\
                               & \multicolumn{1}{l|}{.}         & .            & .            &     & .                & \multicolumn{1}{l|}{.}          & .            \\
                               & \multicolumn{1}{l|}{.}         & .            & .            &     & .                & \multicolumn{1}{l|}{.}          & .            \\
                               & \multicolumn{1}{l|}{$u_r$}     & $n_{r1}$     & $n_{r2}$     & ... & $n_{rq}$         & \multicolumn{1}{l|}{$n_{r(q+1)}$}& $n_{r\bullet}$            \\ \cdashline{2-8}
                               & \multicolumn{1}{l|}{$u_{(r+1)}$} & $n_{(r+1)1}$   & $n_{(r+1)2}$   & ... & $n_{(r+1)q}$               & \multicolumn{1}{l|}{$n_{(r+1)(q+1)}$}   & $n_{(r+1)\bullet}$   \\ \cline{2-8} 
                               & \multicolumn{1}{l|}{Sum}      & $n_{\bullet 1}$     & $n_{\bullet 2}$     & ... & $n_{\bullet q}$         & \multicolumn{1}{l|}{$n_{\bullet(q+1)}$}   & $n_{\bullet \bullet}=n$     
\end{tabular}
\end{table}

According to the operationalization of the similarity of clusterings, the splitting and merging of clusters can be considered either equally (symmetric measure) or differently (asymmetric measure). Therefore, two types of indices are presented in the next section.

\subsection{The Modified Rand Index (MRI)}

When the splitting and merging of clusters have to be considered as factors that reduce the stability (or similarity) of two partitions, the Modified Rand Index (MRI) has to be applied. The Modified Rand Index is defined as the ratio between the number of all possible pairs of units placed in the same or different clusters in both partitions $U$ and $V$ calculated on the contingency table $\mathcal{M}_{U' \times V'}$, and the number of all possible pairs of units from the contingency table $\mathcal{M}_{U' \times V'}$:

\begin{equation}\label{eq:mri}
MRI = \frac{\binom{m}{2} + \sum_{i=1}^r \sum_{j=1}^q n_{ij}^2 - \frac{1}{2} {\Big (} \sum_{i=1}^r n_{i \bullet}^2 + \sum_{j=1}^q n_{\bullet j}^2 {\Big )}}{{n \choose 2}}
\end{equation}

\noindent where $m=\sum_{i=1}^r \sum_{j=1}^q n_{ij}$. Because the splitting and merging of clusters are considered equally (along with the impact of the newcomers and outgoers), the labeling of the first and second partitions is not important. Further, when there are no newcomers and outgoers the Modified Rand Index is equal to the Rand Index. The main properties of the Modified Rand Index are:
\begin{itemize}
  \item it takes a value on the interval between 0 and 1, where a higher value indicates more stable partitions. The value 1 is only possible when there are no newcomers and outgoers. Let us assume that all units from the first partition are classified into their own clusters and no newcomers and outgoers are present. As shown, the value of the Modified Rand Index (or Rand Index) would approach 1 when the number of clusters in the second partition approaches the number of units $n$. However, if some newcomers were present ($n > m$), the value of the Modified Rand Index would approach the limit which is below 1: 

\begin{equation}
\lim_{q \to m} RI = \frac{{m \choose 2}}{{n \choose 2}} < 1
\end{equation}

\noindent similar is true for when outgoers are present and the limit would be even lower if both newcomers and outgoers are present;   
  \item the merging and splitting of clusters results in a lower index value, which can be shown in the same way as for the original Rand Index.
\end{itemize}

\subsection{The Modified Wallace Index 1 and the Modified Wallace Index 2}

While the Rand Index and the Modified Rand Index are both symmetric measures, the Wallace Index and the newly proposed modified Wallace indices are asymmetric measures. This means that the splitting of clusters in the case of the Wallace Index 1 and the Modified Wallace Index 1 lowers the value of the index, but the merging of clusters does not. On the other hand, in the case of the Wallace Index 2 and the Modified Wallace Index 2 merging of clusters lower the value of the index, while the splitting of clusters does not. Consequently, in contrast to the Rand Index and Modified Rand Index, the information about the order of the partitions has to be taken into account. Therefore, the set of units $S$ is considered as a reference set of units (e.g. it is measured at a certain time point, while the set $T$ is measured at a later time point). As in the case of the Modified Rand Index, the newcomers are seen as a factor which lowers the level of stability of clusters from the first partition.

Sometimes, there are situations when one set of units is a subset of another set of units. This means that either there are only newcomers present or only outgoers present in one set of units. Hence, two special cases of Modified Wallace index are also presented.

The Modified Wallace Index 1 is defined as the proportion of all pairs of units that are placed in the same cluster in $U$ and $V$ and all possible pairs of units that are placed in the same clusters in $U'$:

\begin{equation} \label{eq:mwi}
MW1=\frac{\frac{1}{2} \sum_{i=1}^{r} \sum_{j=1}^{q} n_{ij}(n_{ij} - 1)}{\frac{1}{2} \sum_{i=1}^{r+1} \sum_{j=1}^{q+1} n_{ij}(n_{ij} - 1) + \frac{1}{2} (\sum_{i=1}^{r+1} n_{i \bullet}^2 - \sum_{i=1}^{r+1} \sum_{j=1}^{q+1} n_{ij}^2)}
\end{equation}

The Modified Wallace Index 2 is defined as the proportion of all pairs of units that are placed in the same cluster in $U$ and $V$ and all possible pairs of units that are placed in the same clusters in $V'$:

\begin{equation} \label{eq:mwi2}
MW2=\frac{\frac{1}{2} \sum_{i=1}^{r} \sum_{j=1}^{q} n_{ij}(n_{ij} - 1)}{\frac{1}{2} \sum_{i=1}^{r+1} \sum_{j=1}^{q+1} n_{ij}(n_{ij} - 1) + \frac{1}{2} (\sum_{j=1}^{q+1} n_{\bullet j}^2 - \sum_{i=1}^{r+1} \sum_{j=1}^{q+1} n_{ij}^2)}
\end{equation}

\noindent The main properties of the Modified Wallace Index 1 and the Modified Wallace Index 2 are:

\begin{itemize}
  \item it takes a value on the interval between 0 and 1, where a higher value indicates more stable partitions (the value 1 is only possible when there are no newcomers in partition $U'$ and no outgoers in partition $V'$);
  \item in the case of the Modified Wallace Index 1, the merging of clusters does not result in a lower value of the index while the splitting of clusters lowers the value of the index;
  \item in the case of the Modified Wallace Index 2, the splitting of clusters does not lower the value of the index while the merging of clusters lowers value of the index.
\end{itemize}

\subsection{Modified indices considering only outgoers or only newcomers}

It sometimes happens that one set of units is a subset of another. In this case, the sets of units are usually ordered. This means either newcomers are present in the second set of units or outgoers in the first set of units. Hence, four special cases of modified Wallace indices are presented.

When only outgoers are present, the Modified Wallace Index 1 and Modified Wallace Index 2 can be modified by replacing $r+1$ with $r$ in the denominator of Eq.~\ref{eq:mwi} or Eq.~\ref{eq:mwi2}. The indices so modified are called the Modified Wallace Index Outgoers 1 (MWO1) and Modified Wallace Index Outgoers 2 (MWO2). On the other hand, when only newcomers are present, the modified Wallace indices can be simplified by replacing $q+1$ with $q$ in the denominator of Eq.~\ref{eq:mwi} or Eq.~\ref{eq:mwi2}. The indices are then called the Modified Wallace Index Newcomers 1 (MWN1) and Modified Wallace Index Newcomers 2 (MWN2).

When the splitting and merging of clusters must be considered equally, the Modified Rand Index can be used. In order to obtain the value of the Modified Rand Index where only outgoers or only newcomers are present, Eq.~(\ref{eq:mri}) can be used in both cases. This is because the Modified Rand Index is symmetric, which results in the same nominator ($a$ and $d$ are calculated on the $\mathcal{M}_{U' \times V'}$ contingency table neglecting the outgoers and newcomers) and denominator (the number of all possible pairs of units ${n \choose 2}$ where $n$ is the number of units in $S \cup T$) in three possible cases (in the case where both outgoers and newcomers are present, in the case of only outgoers being present, and in the case of only newcomers being present).

\section{Adjustment for chance} \label{correction}

One of the most desirable property of the indices for comparing two partitions is the constant baseline property \cite[p. 2843]{vihn2010} which is, for example, needed to compare the values of indices for different partitions or to test some statistical hypotheses. When the expected value in the case of two random and independent partitions ($Expected~value*$) is not constant (e.g. equal to 0), it can be corrected for chance.

The assumption of a constant value is violated in all proposed indices. This can be shown by a Monte Carlo simulation where the impact of three fixed factors is observed: (i) the number of clusters in the partition $U$ (from 8 to 24 by 2); (ii) the number of clusters in the partition $V$ (from 8 to 24 by 2); and (iii) the total number of units in $U'$ or $V'$ (depends on the index) (from 100 to 220 by 20). There are 300 random and independent partitions for each combination of factors ($9 \times 9 \times 7$ design) and so the potential value greater than 0 must be caused by agreement due to chance.

If clusters of an equal size at one point in time are assumed (including newcomers and outgoers), which is very unrealistic, then it can be shown in Fig.~2 that the expected values of all presented indices are not constant. \citet[p. 2847]{vihn2010} highlighted that "adjustment for chance for information theoretic measures is mostly needed when the number of data items is relatively small compared to the number of clusters". As noted before, the explanation is valid in the case of the Rand Index (not included in Fig.~2) and the Modified Rand Index where the expected value in the case of two random and independent partitions is increasing with the number of clusters, while in the cases of the Modified Wallace Index 1 the expected value in the case of two random partitions is decreasing when the number of clusters in the second partition is increasing and simmilarly, in the case of the Modified Wallace Index 2 the expected value in the case of two random and independent partitions is decreasing when the number of clusters in the first partition is increasing (Fig.~2). 

\begin{figure}[ht!]
  \setkeys{Gin}{width=1\textwidth}
\includegraphics[angle =0]{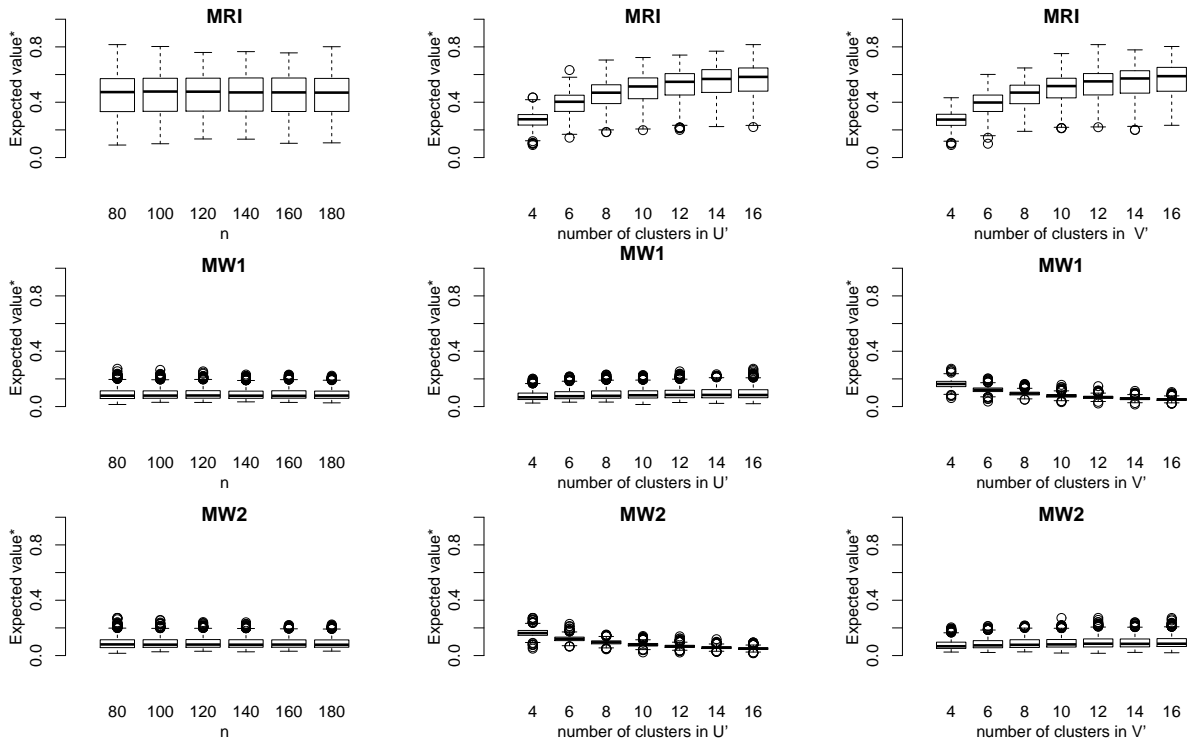}
  \caption{Expected values in the case of two random and independent partitions of the Modified Rand Index (MRI), Modified Wallace Index (MW), Modified Wallace Index 1 (MW1) and Modified Wallace Index 2 (MW2) regarding the values of factors}
  \label{pricakovane}
\end{figure}

This effect arises because the merging and splitting of clusters do not have the same impact on the value of the Wallace indices and the Modified Wallace Index 1 and the Modified Wallace Index 2. In the case of the Wallace Index 1 and the Modified Wallace Index 1, a higher number of clusters in $V$ (in comparison with $U$) increases the probability of splitting, which results in a lower value of the measure, but a higher number of clusters in $U$ (in comparison with $V$) increases the probability of the merging of clusters which does not affect the value of the measure. As a result, it seems that the impact of the number of clusters in $U$ on the expected value of the measure is almost constant. On the other hand, in the case of the Wallace Index 2 and the Modified Wallace Index 2 a higher number of clusters in partition $U$ (in comparison with $V$) is associated with a higher probability of merging of clusters which lowers the value of mentioned indices.

As used by many researchers (e.g., \cite{cohen1960}, \cite{hubertandarabie1985}, \cite{morlini2012}), the general form of the index corrected by chance is as follows:

\begin{equation} \label{eq:expected}
Adjusted~Index=\frac{Index-Expected~value*}{Maximum~Index-Expected~value*}
\end{equation}

\noindent where the upper bound of the adjusted index is 1 and the expected value in the case of two random and independent partitions is 0. Much attention has been paid to the question of how to obtain (or estimate) the $Expected~value*$. Assuming the approximation:

\begin{equation} \label{eq:ari_manda}
E{\Big (}\sum_{i=1}^I \sum_{j=1}^J n_{ij}^2 {\Big )} \approx \sum_{i=1}^I \sum_{j=1}^J \frac{n_{i \bullet}^2 n_{\bullet j}^2}{n^2}
\end{equation}

\noindent  \cite{morey1984measurement} obtained an asymptotic correction by chance for the Rand Index based on an asymptotic multinomial distribution. \citet[p. 200]{hubertandarabie1985} argued that "the difference between the adjusted indices is not necessarily small, depending on the sizes of the object sets and associated partitions being compared". Later, \citet[p. 308]{albatineh2006} showed that the differences between the mentioned methods are negligible when the number of objects for clustering is large enough. However, \cite{hubertandarabie1985} corrected the Rand Index for chance on the assumption of a generalized hypergeometric distribution:

\begin{equation} \label{eq:ari}
E{\Big (}\sum_{i=1}^I \sum_{j=1}^J n_{ij}^2 {\Big )} = \frac{\sum_{i=1}^I \sum_{j=1}^J n_{i \bullet}^2 n_{\bullet j}^2}{n(n-1)} +
\frac{n^2 - (\sum_{i=1}^I n_{i \bullet}^2 + \sum_{j=1}^J n_{j\bullet}^2)}{m-1}
\end{equation}

The $Expected~value*$ for $W1$ and simmilarly for $W2$ can be defined by Simpson's index of diversity (SID) of the $V$ (for W1) and $U$ (for W2) partitions  \citep{simpson1949, pinto2001}:

\begin{equation}\label{eq:simpson}
SID_U = \frac{\sum_{i=1}^r n_{i\bullet} (n_{i\bullet} - 1)} {n(n-1)} \qquad\text{and}\qquad SID_V = \frac{\sum_{j=1}^q n_{\bullet j} (n_{\bullet j} - 1)} {n(n-1)}
\end{equation}

The common methods to estimate the $Expected~value*$ assume that the contingency table $\mathcal{M}$ is constructed from a generalized hypergeometric distribution, i.e. the partitions are picked at random, subjected  to have the original number of clusters and units in each cluster \citep[p. 197]{hubertandarabie1985}. The expected values of the proposed modified indices in the case of two random partitions $U$ (or $U'$) and $V$ (or $V'$) cannot be determined by the mentioned methods on the assumption of a generalized hypergeometric distribution. Due to the already mentioned fact that the marginals of the contingency tables, excluding the newcomers and outgoers, are not constant, especially when there are high values of the non-adjusted measure. In this case, the difference between the expected marginal frequencies of the $U' \times V'$ contingency table (excluding the newcomers and outgoers), $U' \times V$ contingency table (excluding the newcomers) or the $U \times V'$ contingency table (excluding the outgoers) in the case of two random and independent partitions and empirical marginal frequencies is usually greater.

\begin{algorithm}[H]
\caption{Estimating the $Expected~value*$ by simulations}
\label{algecsi}
\begin{algorithmic}[1]
\State import partition $U$ or $U'$                     
\State import partition $V$ or $V'$                     
\State $R \gets NULL$
\State $k \gets number~of~repetitions$
\For{$i~in~1:k$}
    \State $U* \gets$ randomly permute the units of $U$ or $U'$
    \State $V* \gets$ randomly permute the units of $V$ or $V'$
    \State $R_i \gets$ calculate the value of the index from~$U*$~and~$V*$ 
\EndFor
\State \Return $mean(R)$
\end{algorithmic}
\end{algorithm}

\begin{figure}[H]
  \setkeys{Gin}{width=1\textwidth}
\includegraphics[angle =0]{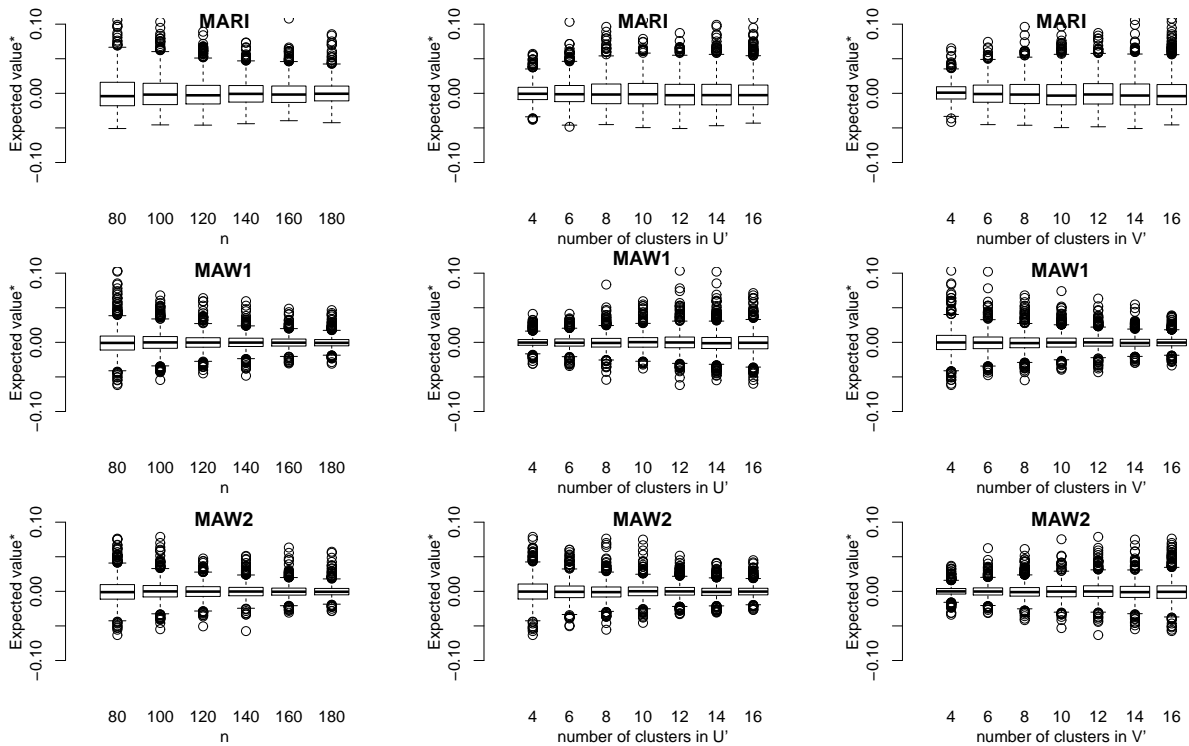}
  \caption{Expected values of the Modified Adjusted Rand Index (MARI), Modified Adjusted Wallace Index 1 (MAW1) and  Modified Adjusted Wallace Index 2 (MAW2) regarding the values of factors}
  \label{expected_adjusted}
\end{figure}

One approach to estimating the $Expected~value*$ can be based on simulations\footnote{Due to the fact that there is no analytical expression of the confidence interval proposed for all presented indices except for W1 \citep{pinto2001} which is valid in some conditions, the confidence intervals are usually obtained by simulations. Two of the most common non-parametric estimates of confidence intervals are Jackknife \citep{quenouille1949, tukey1958} and Bootstrap \citep{efron1979}. While some simulation studies do not show the clear superiority of Jackknife or Bootstrap but the impact of sample variability and size on the suitable oversampling method to be applied \citep{smith1986, hellmann1999}, some other evaluation studies of the mentioned methods for defining confidence intervals for pairwise agreement measures show that the Jackknife method performs better than the Bootstrap one, especially when the agreement between the partitions is low \citep{severiano2011}.}. Our algorithm~\ref{algecsi} is similar to the algorithm of the non-parametric permutation test \citep{berry2014}. The simulation is based on empirical data. First, the partitions $U$ or $U'$ and $V$ or $V'$ have to be considered. Then, the units of each partition have to be randomly and independently permuted. In the next step, the value of the index has to be calculated based on these permuted partitions, and the value has to be added to vector $R$. This procedure has to be repeated many times and, after $k$-repetitions, the mean of vector $R$ presents the estimate of the $Expected~value*$.

The expected value calculated by the proposed algorithm takes into account all characteristics of the sets of units and partitions (the number of clusters in each partition, the number of units, and the number of units in each cluster, etc.). By considering the $Expected~value*$ in Eq.~(\ref{eq:expected}) the adjusted indices are obtained. In the case where the non-adjusted value of an index equals 1, the adjusted value of the index is also 1, but in all other cases the adjusted values are lower than the non-adjusted values. In the case of two random and independent partitions, the expected value is constant and equal to 0 (Fig~\ref{expected_adjusted}).

\section{Examples}

The characteristics of the original Rand and Wallace indices and the proposed indices are presented by selected basic examples of misclassifications of units in clusters of the first and second partitions. Use of the presented indices is also illustrated on the empirical data.

\subsection{Selected basic examples of misclassifications of units}

Some examples of misclassifications of units in clusters of the first and second partitions are selected to illustrate the presented indices' properties. Each example in Fig.~4 has two parts. The black rectangles on the top part of each example represent the clusters in the first partition while the black rectangles on the bottom represent the clusters in the second, respectively. The rectangles on the bottom, colored grey, represent the cluster of outgoers and the rectangles on the top, colored grey, represent the cluster of newcomers. The edges connecting the clusters represent the stability of the clusters of the two partitions.

\begin{figure}[H]
    \label{mri}

    \begin{subfigure}[b]{0.24\textwidth}
        \includegraphics[width=\textwidth]{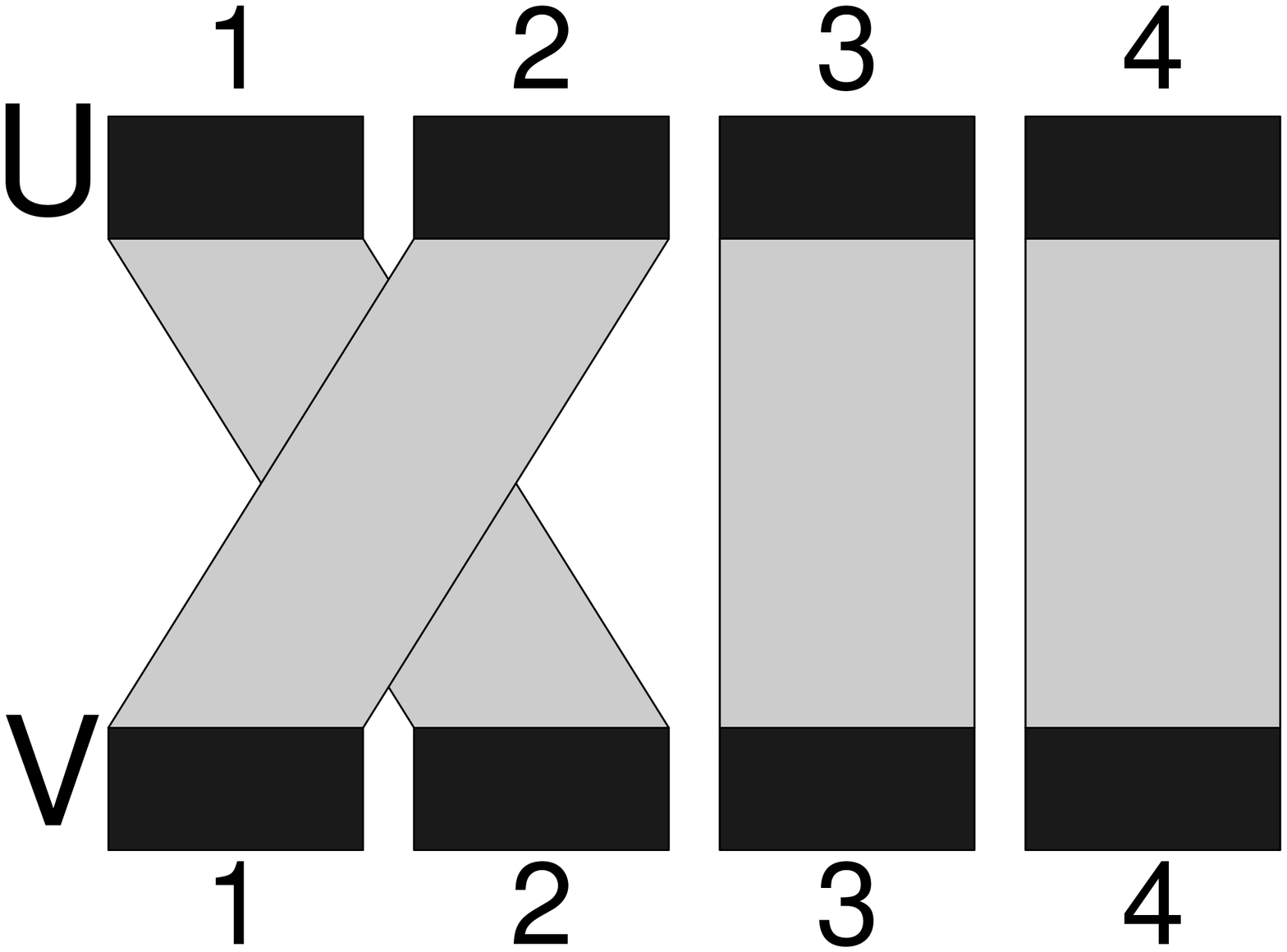}
        \caption{identical}
        \label{a}
    \end{subfigure}%
    \begin{subfigure}[b]{0.24\textwidth}
        \includegraphics[width=\textwidth]{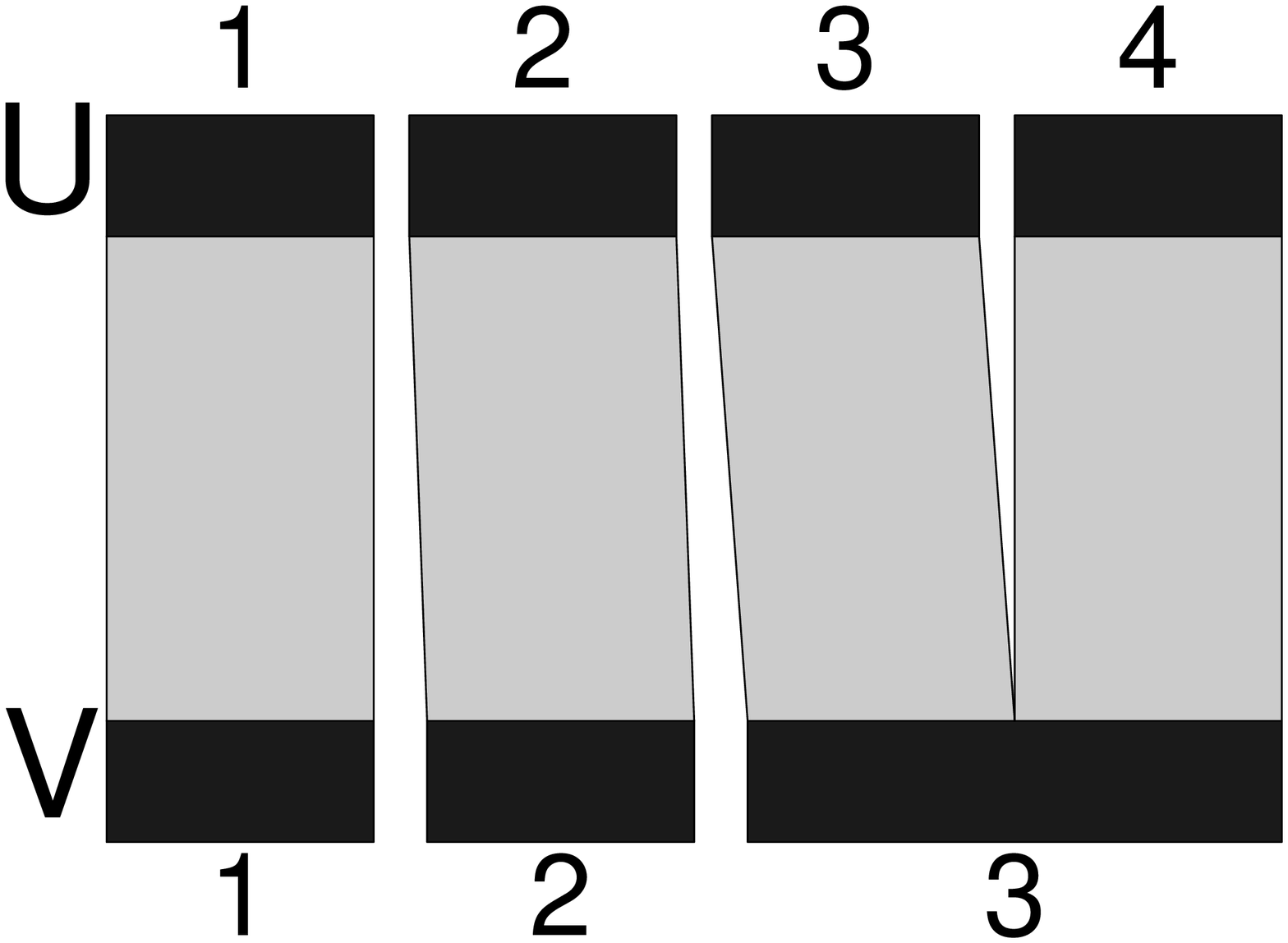}
        \caption{merging}
        \label{b}
    \end{subfigure}
    \begin{subfigure}[b]{0.24\textwidth}
        \includegraphics[width=\textwidth]{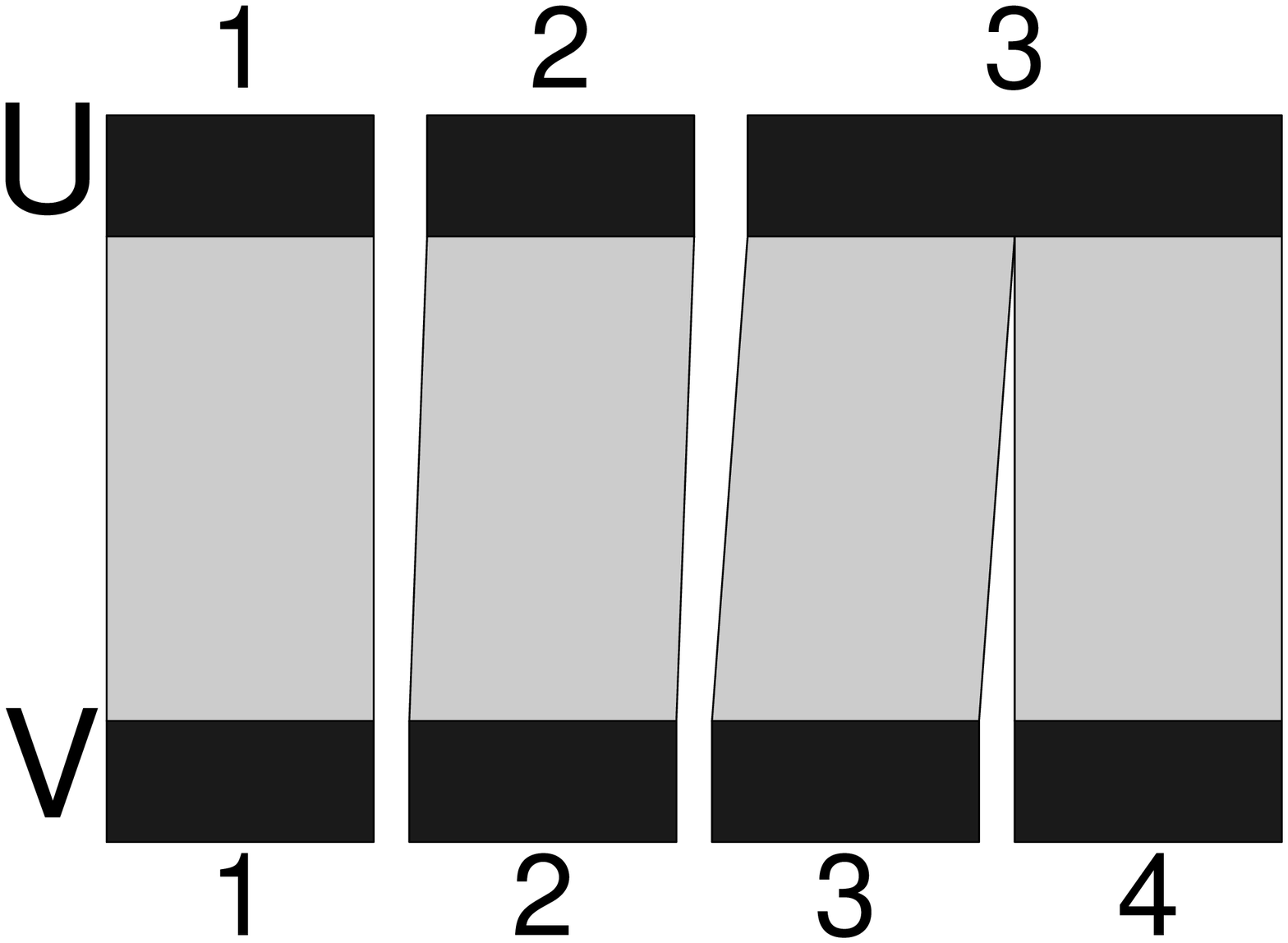}
        \caption{splitting}
        \label{c}
    \end{subfigure}
    \begin{subfigure}[b]{0.24\textwidth}
        \includegraphics[width=\textwidth]{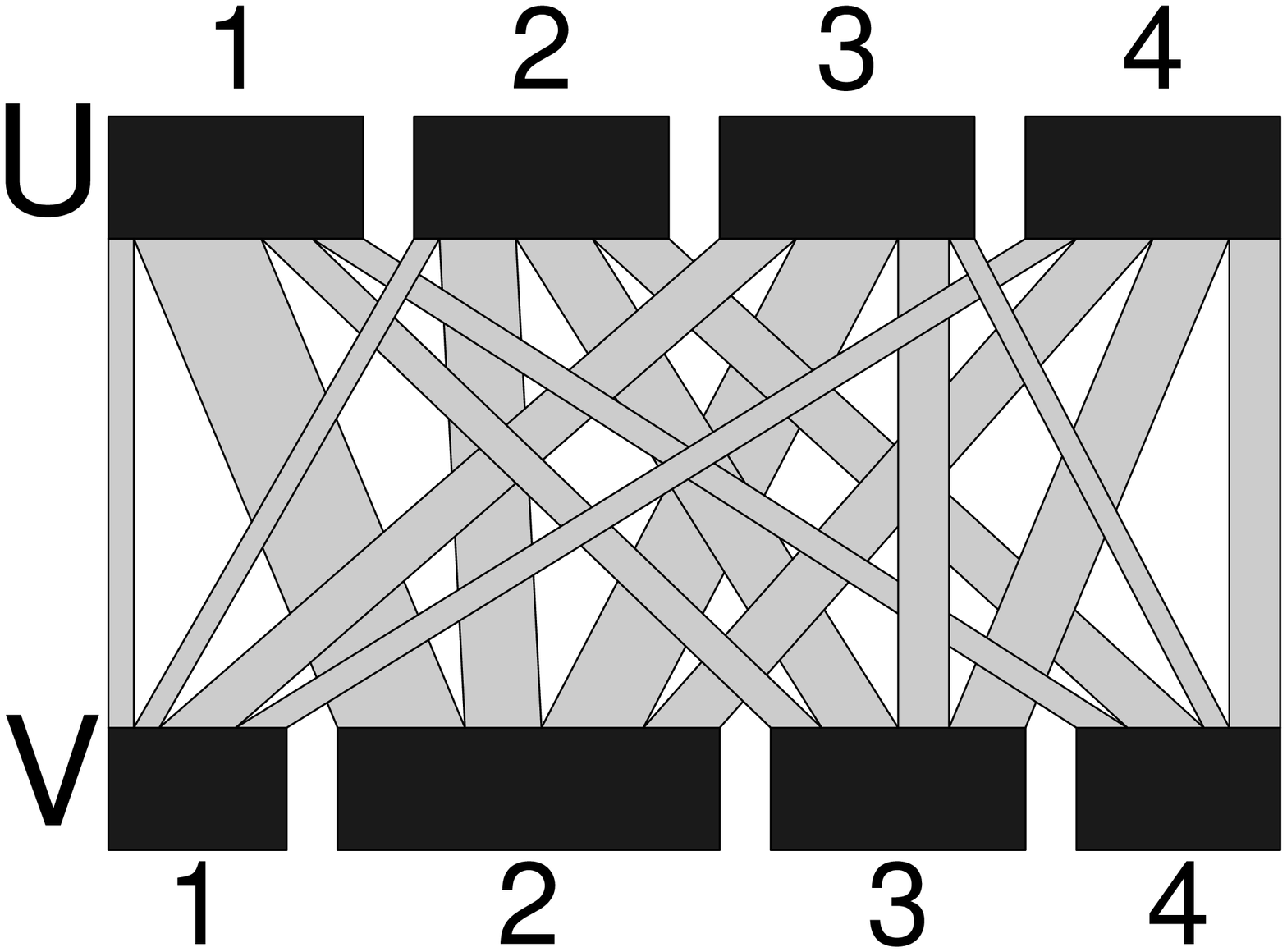}
        \caption{random}
        \label{d}
    \end{subfigure}
    
    \begin{subfigure}[b]{0.24\textwidth}
        \includegraphics[width=\textwidth]{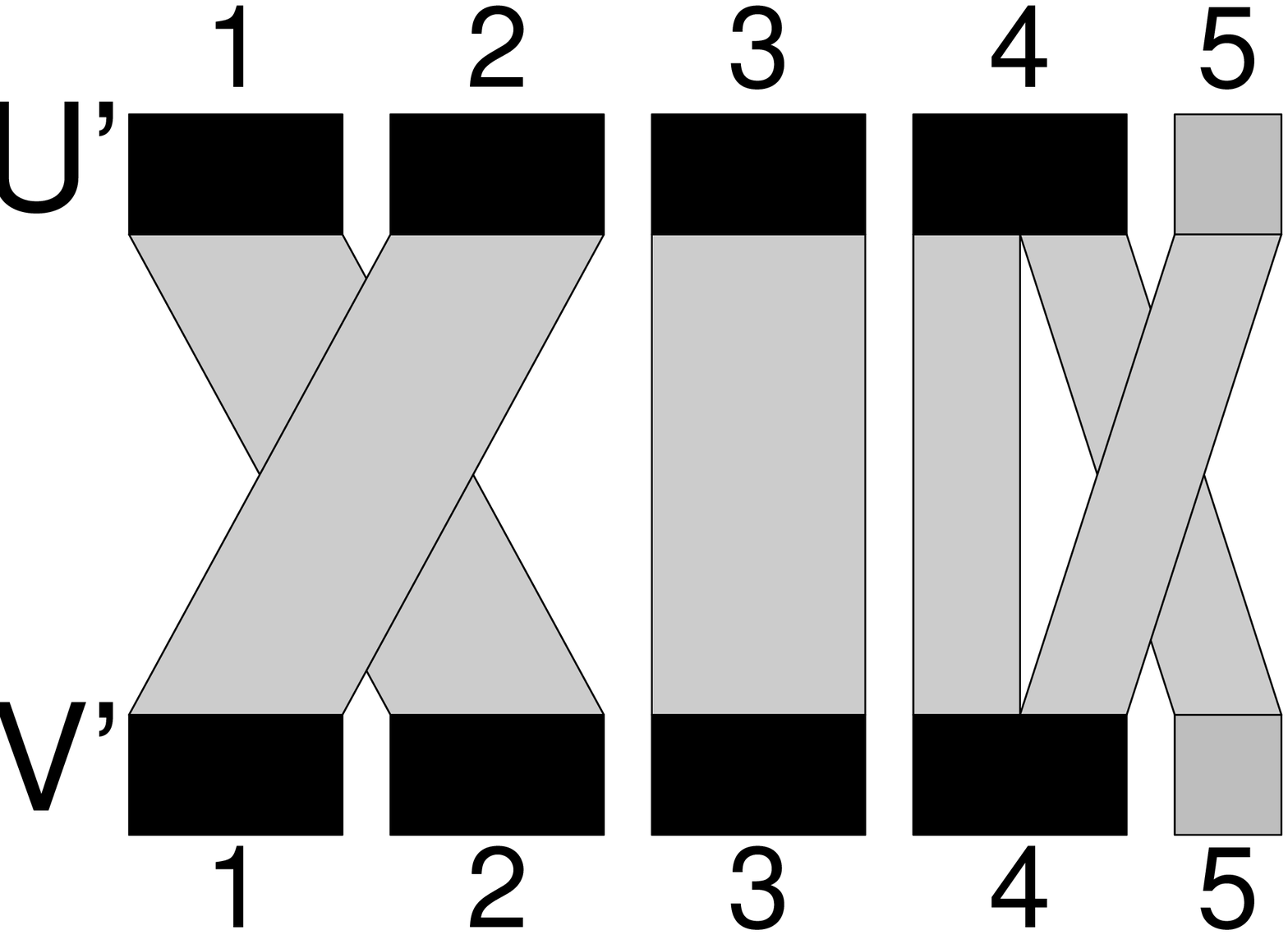}
        \caption{quasi-identical}
        \label{a}
    \end{subfigure}%
    \begin{subfigure}[b]{0.24\textwidth}
        \includegraphics[width=\textwidth]{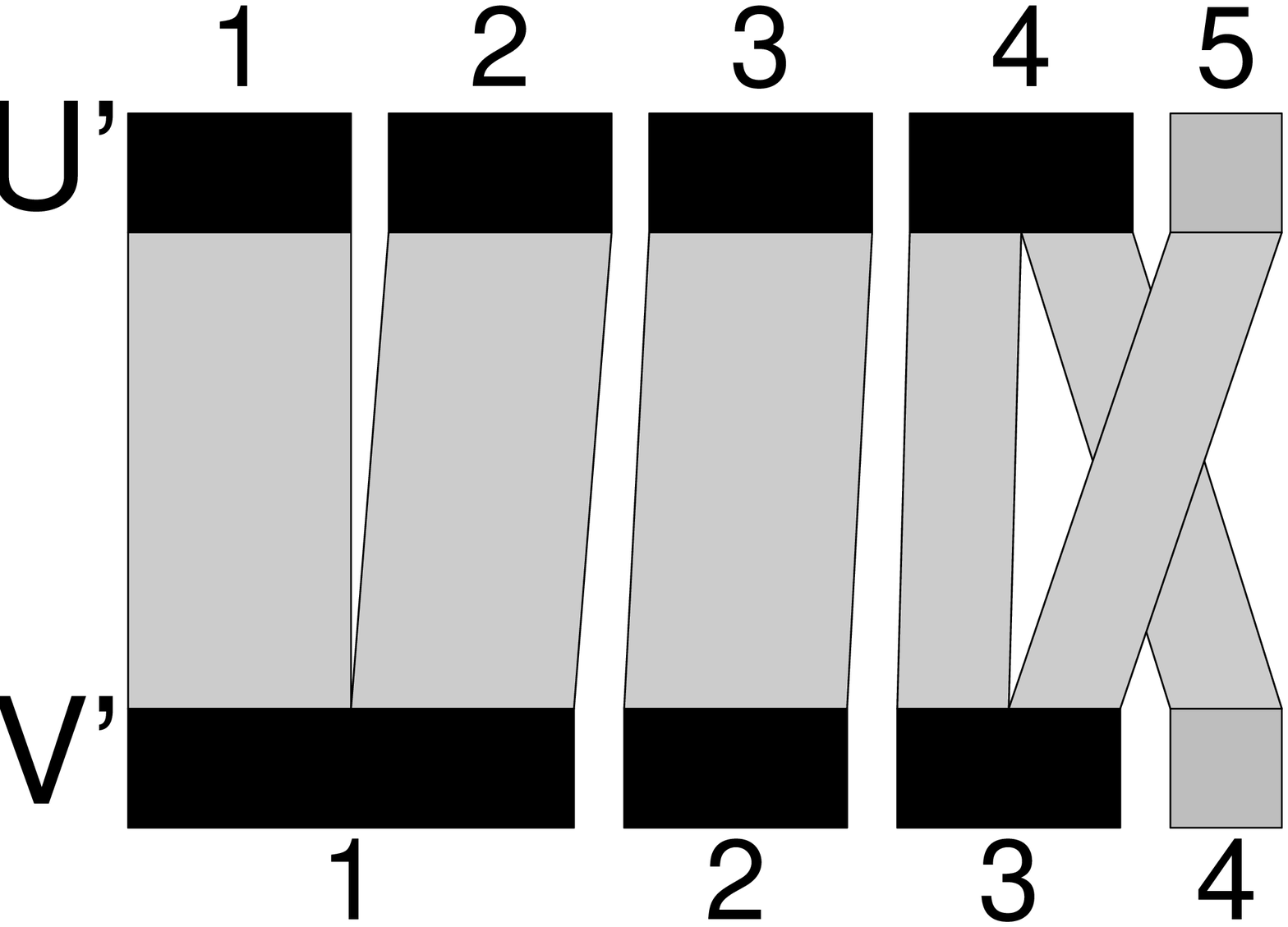}
        \caption{merging}
        \label{b}
    \end{subfigure}
    \begin{subfigure}[b]{0.24\textwidth}
        \includegraphics[width=\textwidth]{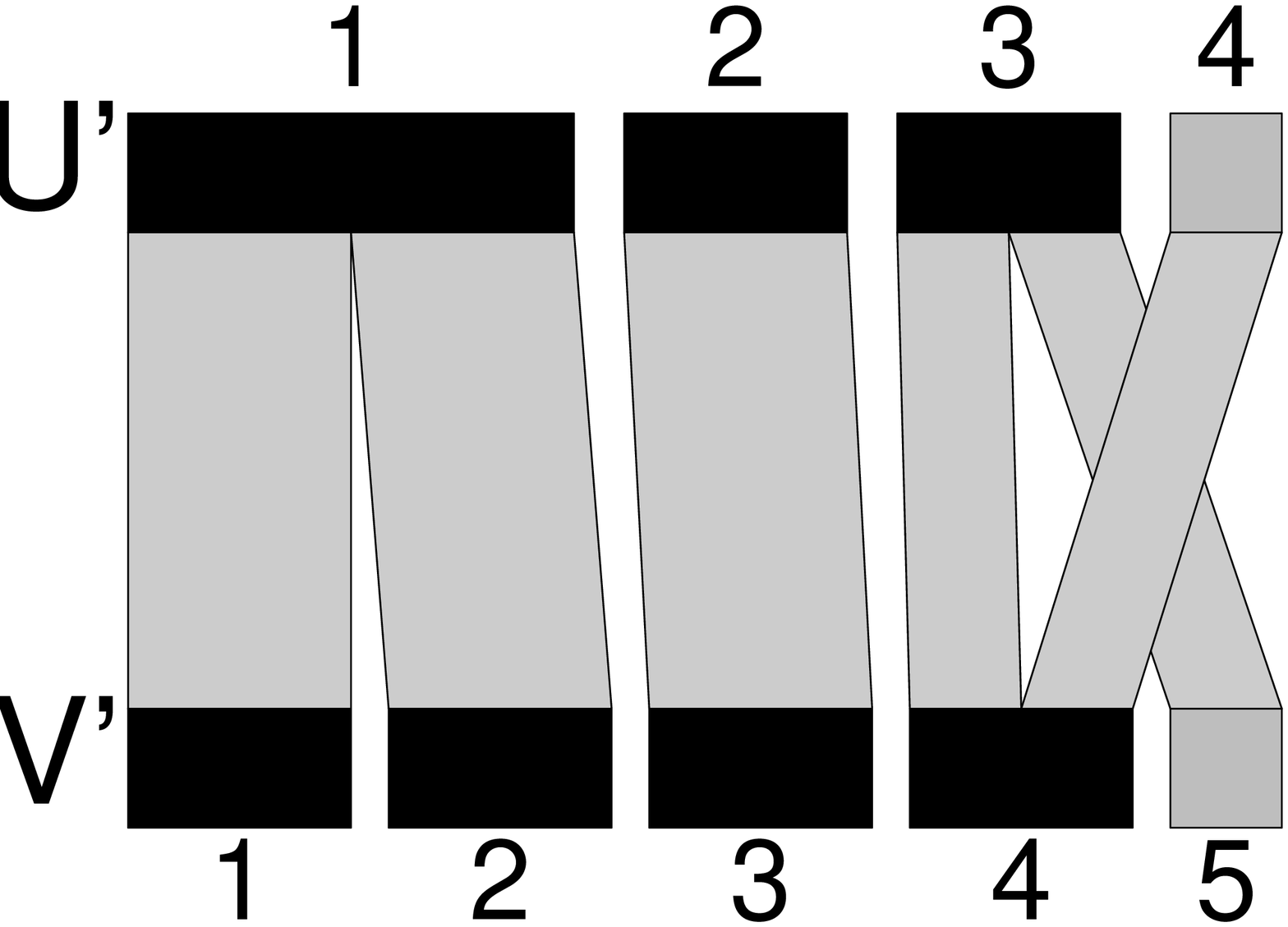}
        \caption{splitting}
        \label{c}
    \end{subfigure}
    \begin{subfigure}[b]{0.24\textwidth}
        \includegraphics[width=\textwidth]{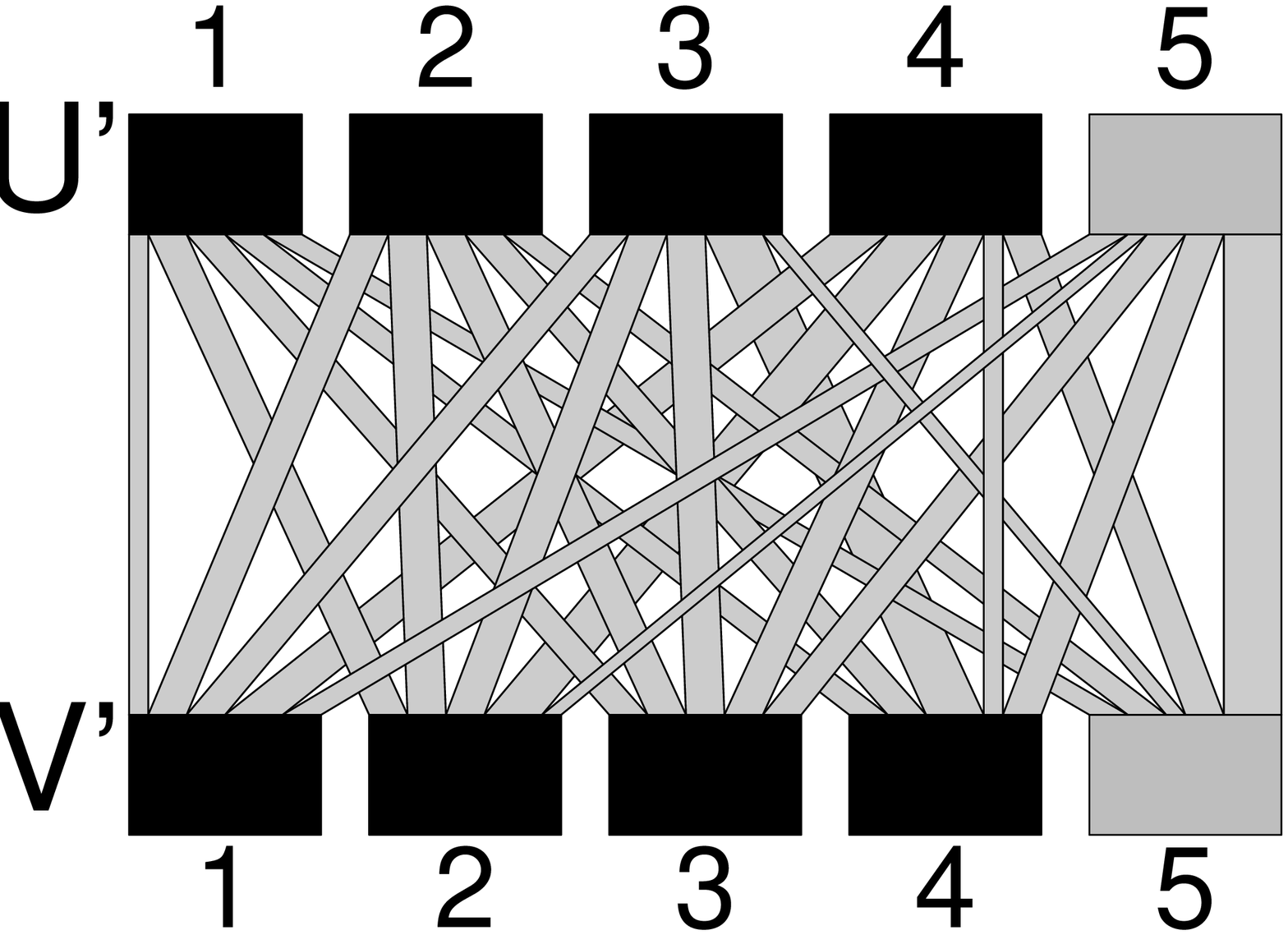}
        \caption{random}
        \label{d}
    \end{subfigure}
    
    \begin{subfigure}[b]{0.24\textwidth}
        \includegraphics[width=\textwidth]{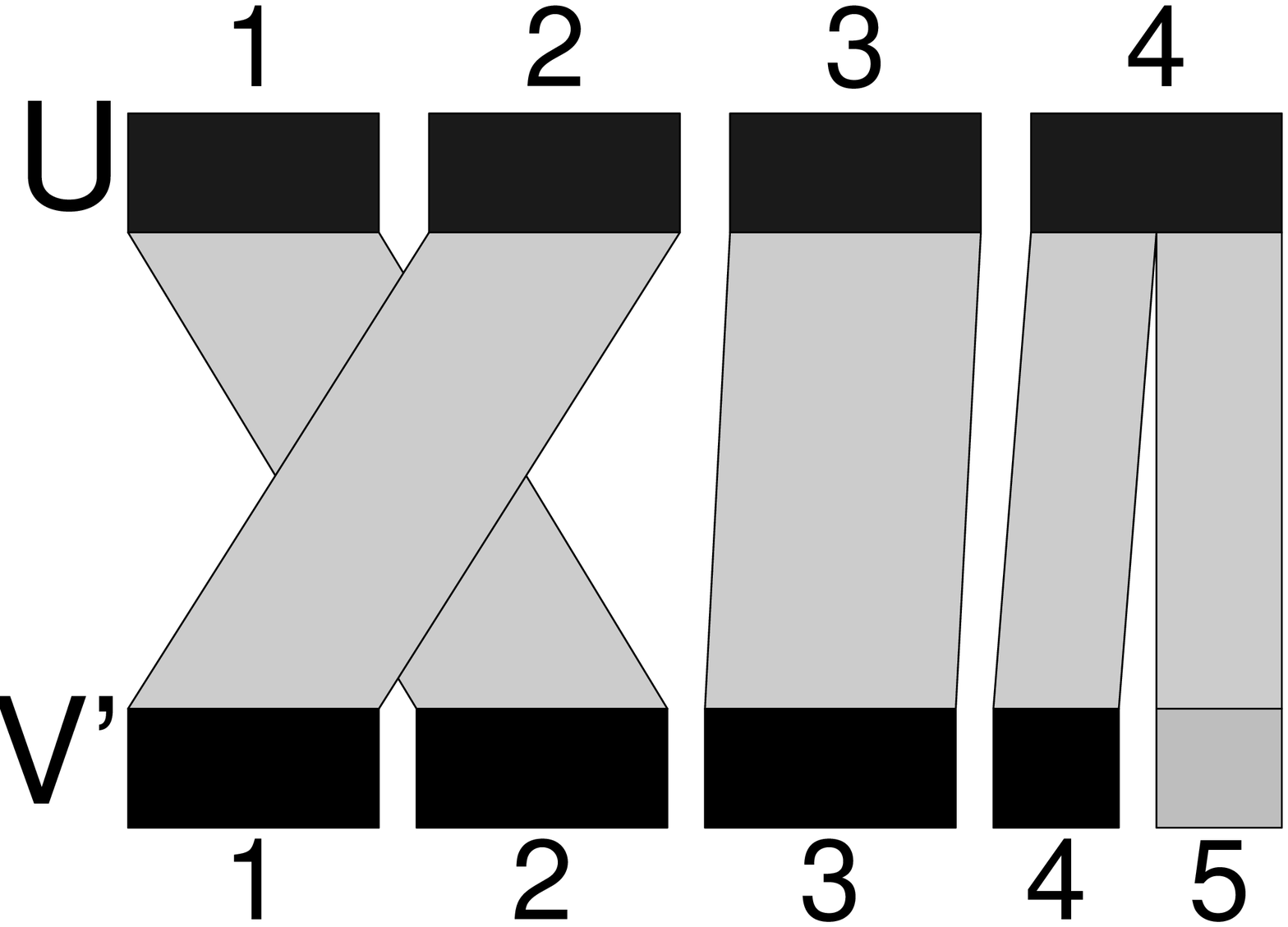}
        \caption{quasi-identical}
        \label{e}
    \end{subfigure}%
    \begin{subfigure}[b]{0.24\textwidth}
        \includegraphics[width=\textwidth]{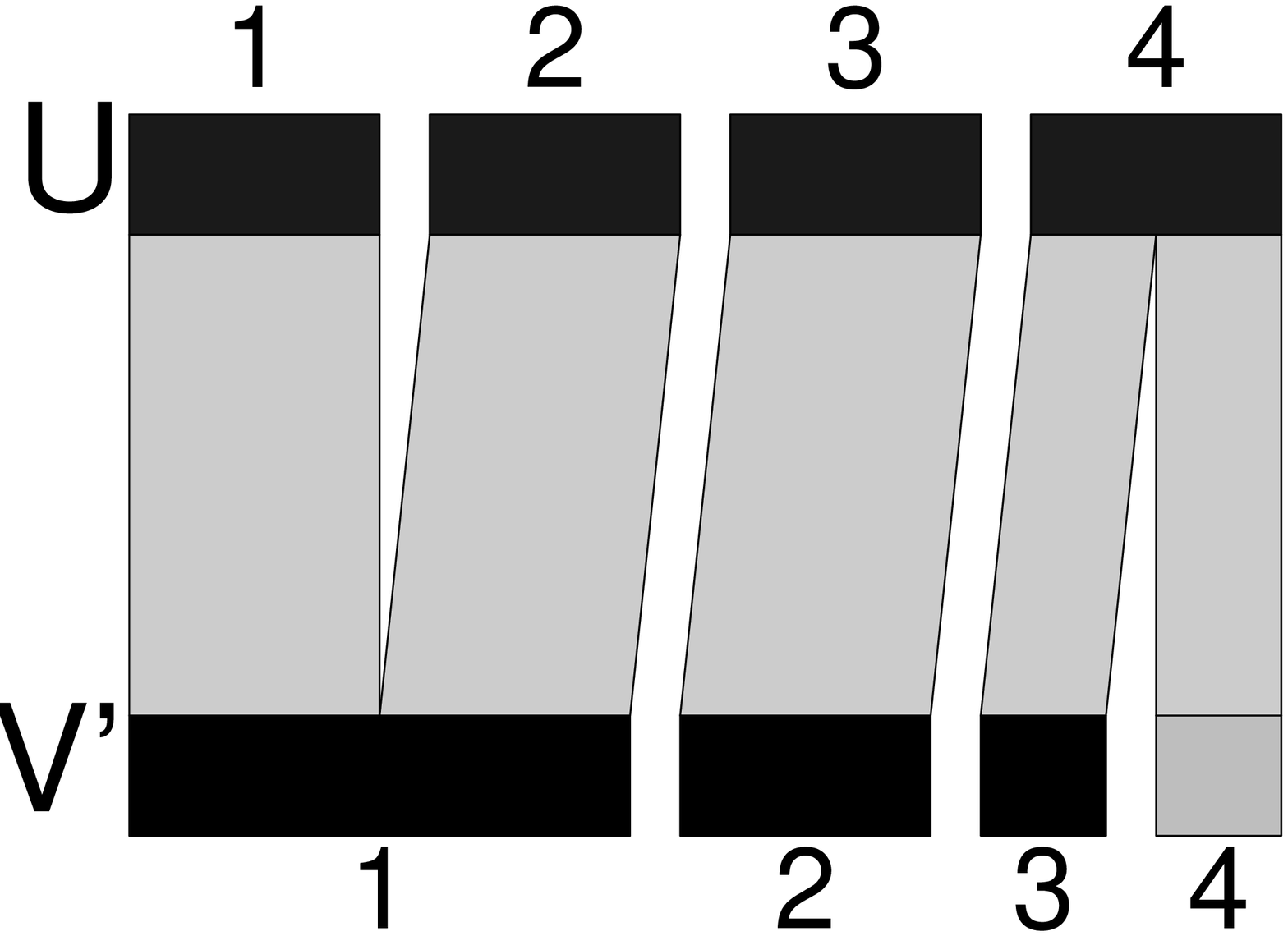}
        \caption{merging}
        \label{f}
    \end{subfigure}
    \begin{subfigure}[b]{0.24\textwidth}
        \includegraphics[width=\textwidth]{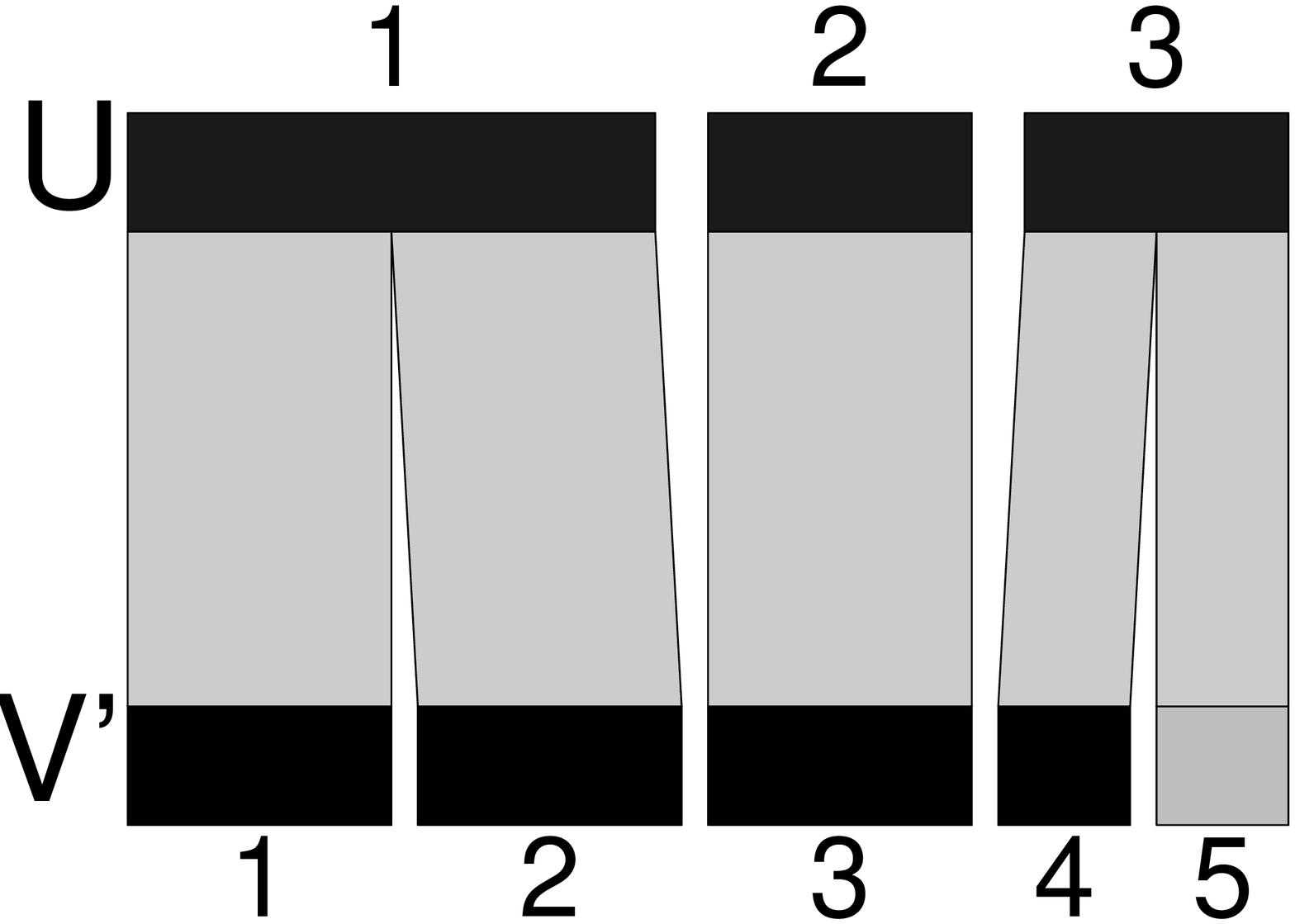}
        \caption{splitting}
        \label{g}
    \end{subfigure}
    \begin{subfigure}[b]{0.24\textwidth}
        \includegraphics[width=\textwidth]{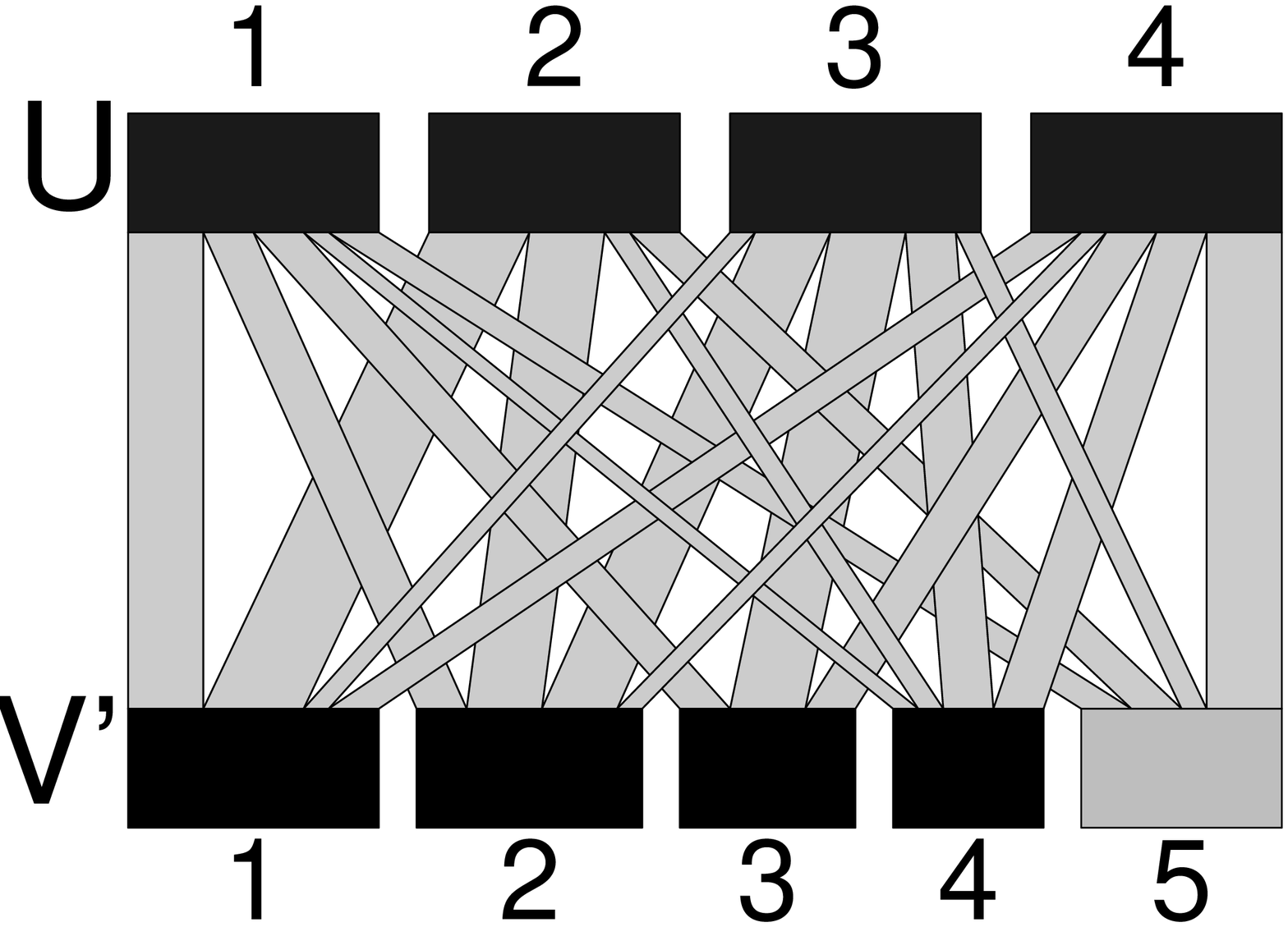}
        \caption{random}
        \label{h}
    \end{subfigure}
    
        \begin{subfigure}[b]{0.24\textwidth}
        \includegraphics[width=\textwidth]{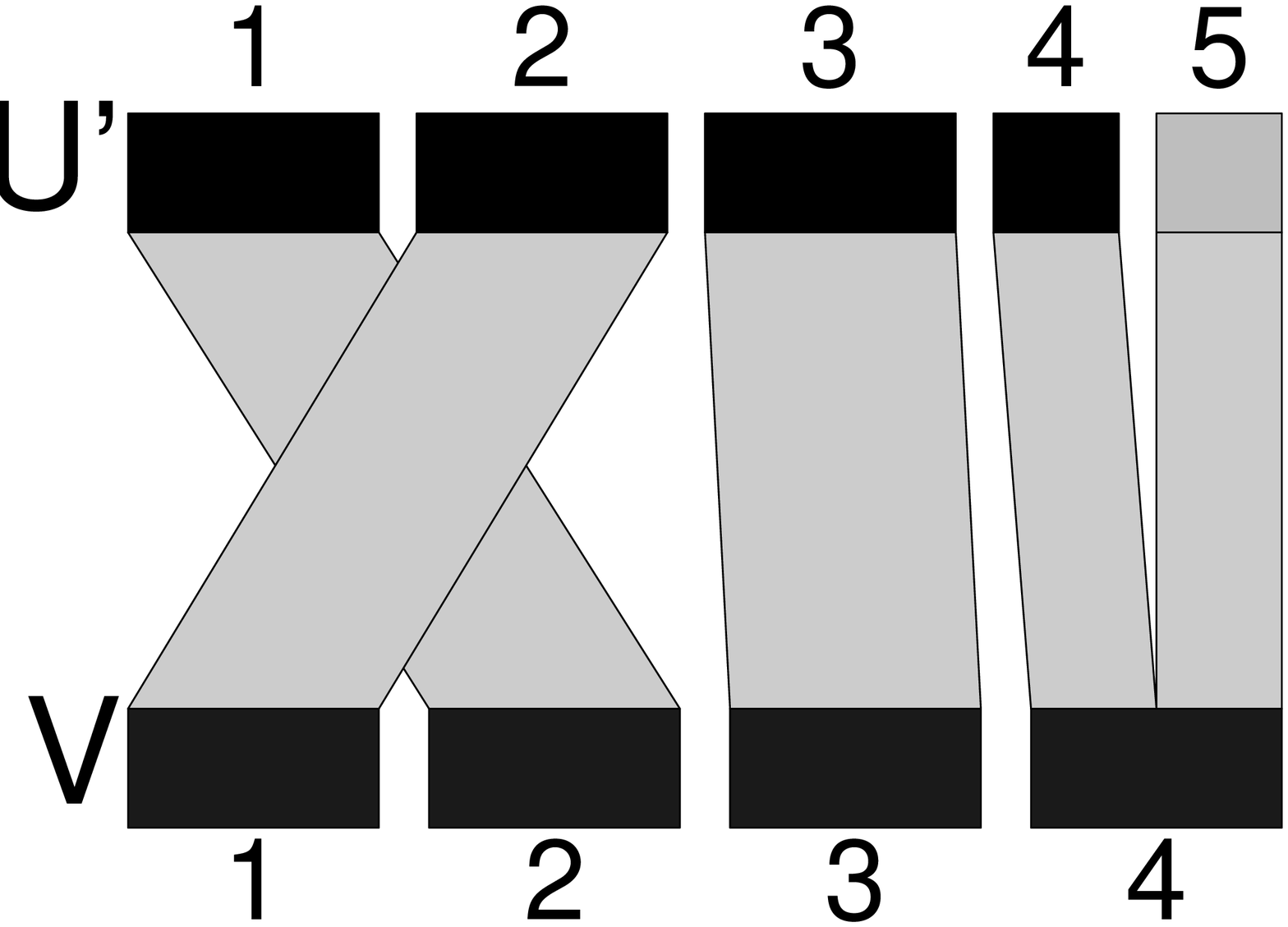}
        \caption{quasi-identical}
        \label{i}
    \end{subfigure}%
    \begin{subfigure}[b]{0.24\textwidth}
        \includegraphics[width=\textwidth]{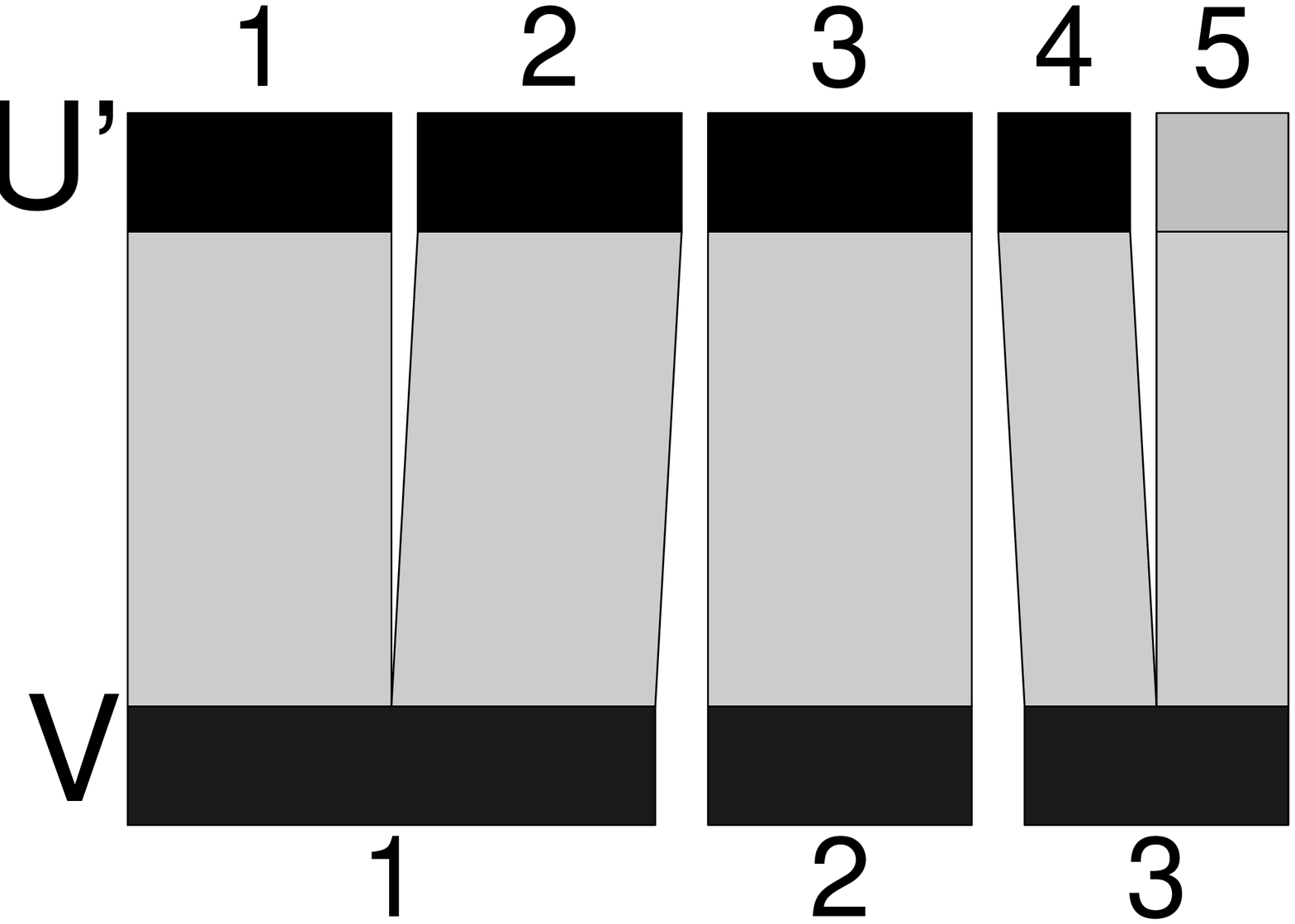}
        \caption{merging}
        \label{j}
    \end{subfigure}
    \begin{subfigure}[b]{0.24\textwidth}
        \includegraphics[width=\textwidth]{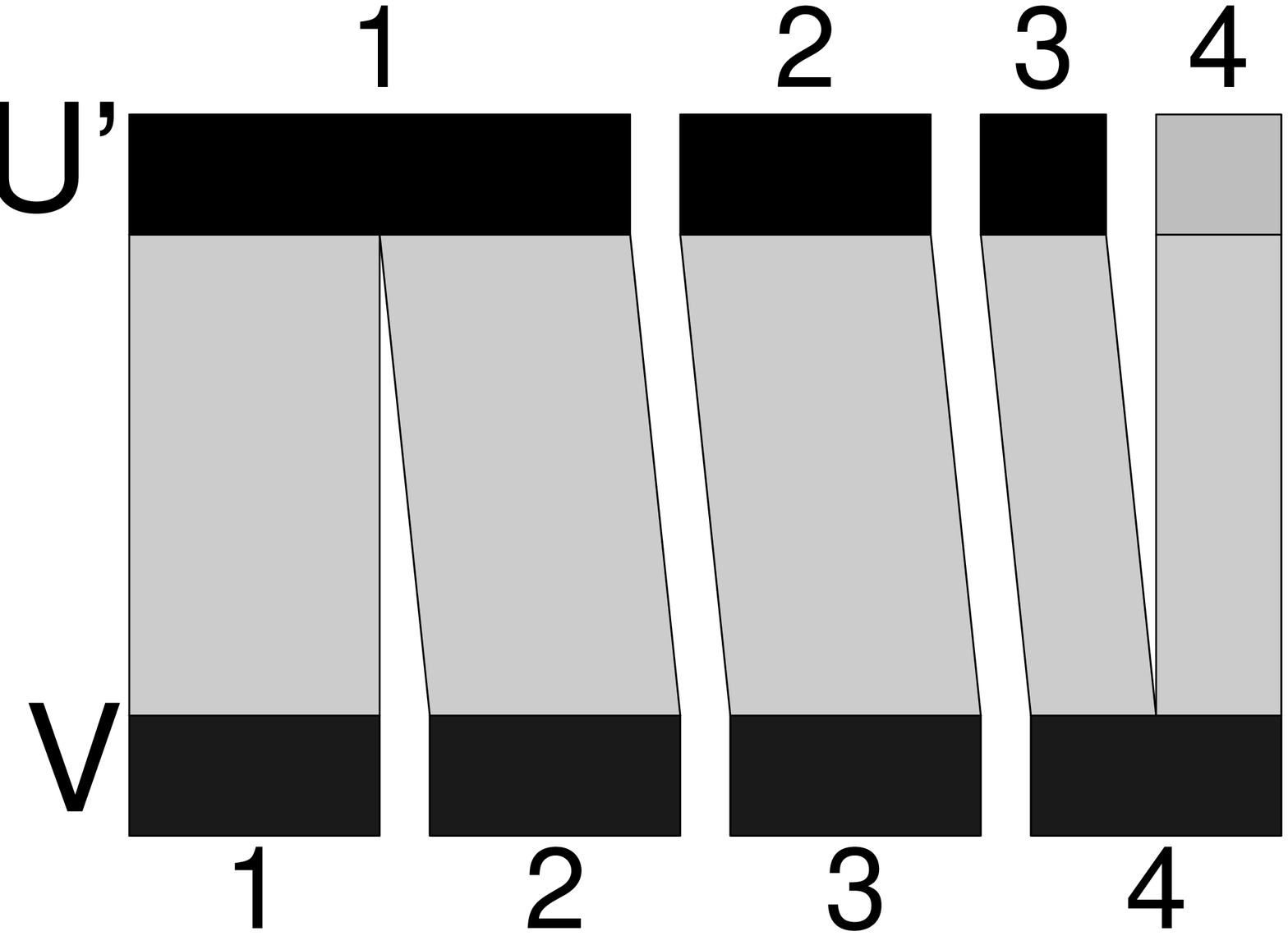}
        \caption{splitting}
        \label{k}
    \end{subfigure}
    \begin{subfigure}[b]{0.24\textwidth}
        \includegraphics[width=\textwidth]{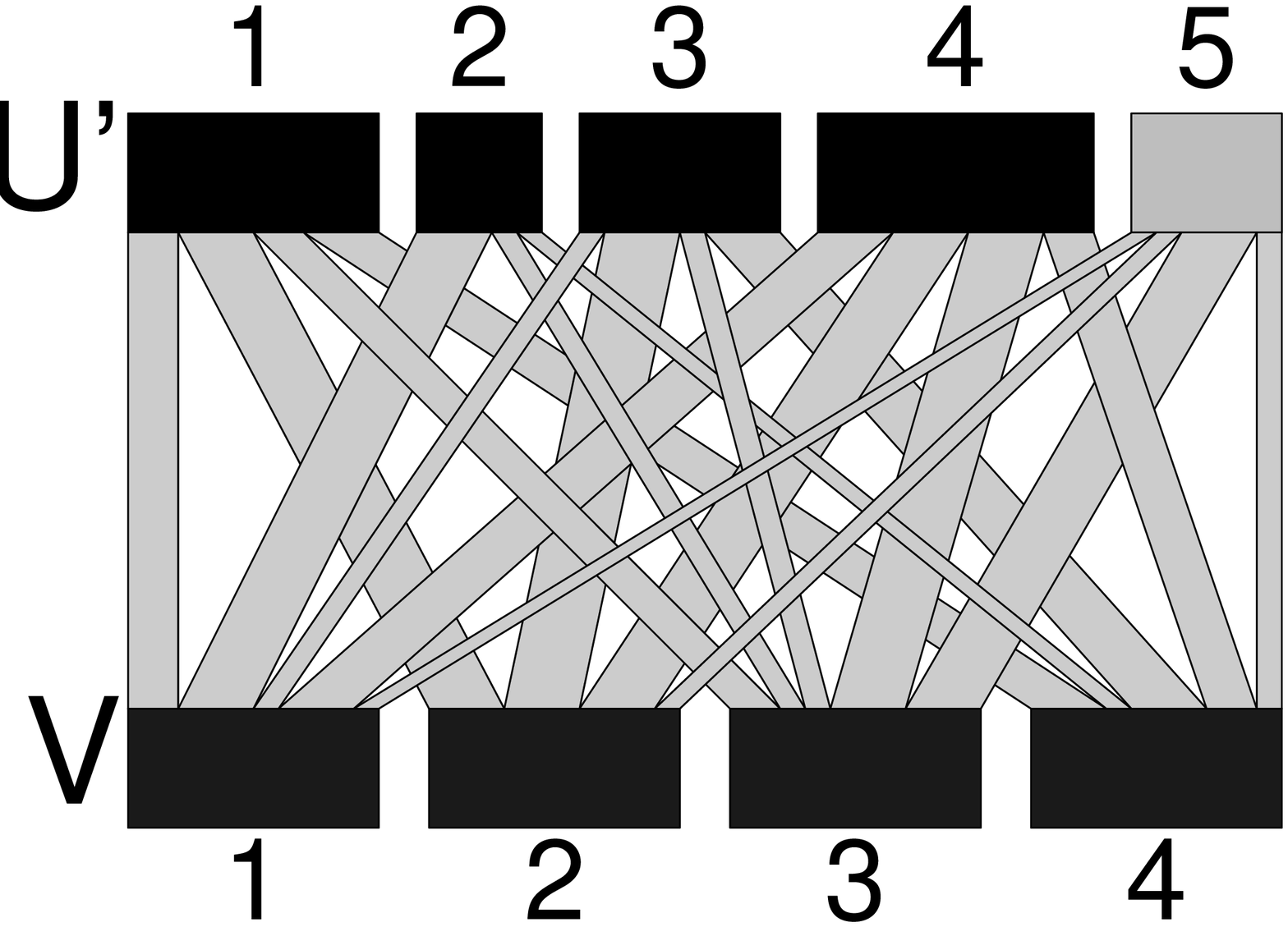}
        \caption{random}
        \label{l}
    \end{subfigure}
        \caption{Visualization of $\mathcal{M}_{U \times V}$, visualisation of $\mathcal{M}_{U' \times V'}$ (the last cluster in partition $U'$ consists of newcomers and the last cluster in partition $V'$ consists of outgoers) and visualization of $\mathcal{M}_{U \times V'}$ (the last cluster in partition $V'$ consists of outgoers) and visualization of $\mathcal{M}_{U' \times V}$ (the last cluster in partition $U'$ consists of newcomers)}
          \label{prehodi_mri}
\end{figure}

Example (a) shows two partitions $U$ and $V$ that are the same. The only difference between $U$ and $V$ is that the first cluster in $U$ is equal to the second cluster in $V$. In example (b), the merging of clusters is visualized while the splitting of clusters is visualized in example (c). Example (d) illustrates two random and independent partitions. The examples in the following rows in Fig.~4 are the same as the examples in the first row, with the only differences being that the cluster of newcomers and/or outgoings (grey colored) is added to each example (because of these, the examples (e), (i) and (m) are called "quasi-identical").

\begin{table}
\centering 
\caption{The values of the Modified Rand Index (the general ones and the special cases) and the Modified Wallace Index (the general ones and the special cases)}
\label{values}
\footnotesize
\begin{tabular}{@{}lllll@{}}
\hline
\begin{tabular}[c]{@{}l@{}}\textbf{the existance of newcomers}\\ \textbf{and outgoings}\end{tabular}  & 
\multicolumn{1}{l}{\begin{tabular}[c]{@{}l@{}}\textbf{identical or} \\ \textbf{quasi-identical}\end{tabular}}  &\textbf{ merging}  &\textbf{ splitting}    & \textbf{random}     \\ \hline \hline   
\textbf{no newcomers or outgoers} & example (a) & example (b) & example (c) & example (d) \\ \hline
Adjusted Rand Index                       & $1.00$            & $0.70$            & $0.70$            & $-0.05$       \\
Adjusted Fowlkes and Mallows Index        & $1.00$            & $0.72$            & $0.72$            & $-0.04$       \\
Adjusted Wallace Index 1                  & $1.00$            & $1.00$            & $0.54$            & $-0.05$       \\
Adjusted Wallace Index 2                  & $1.00$            & $0.54$            & $1.00$            & $-0.05$       \\ \hline
\textbf{both newcomers and outgoers}   & example (e) & example (f) & example (g) & example (h) \\ \hline
M. Adjusted Rand Index                    & $0.34$       & $0.23$       & $0.23$       & $~~~0.00$       \\  
M. Adjusted Wallace Index 1               & $0.71$       & $0.68$       & $0.40$       & $-0.03$       \\  
M. Adjusted Wallace Index 2               & $0.71$       & $0.42$       & $0.69$       & $-0.03$       \\  \hline
\textbf{only outgoers}      & example (i) & example (j) & example (k) & example (l) \\ \hline
M. Adjusted Rand Index                    & $0.54$       & $0.37$       & $0.35$       & $-0.01$       \\                              
M. Adjusted Wallace Index Outgoers 1      & $0.76$       & $0.72$       & $0.41$       & $-0.02$      \\ 
M. Adjusted Wallace Index Outgoers 2      & $0.92$       & $0.45$       & $0.90$       & $-0.03$       \\  \hline
\textbf{only newcomers}       & example (m) & example (n) & example (o) & example (p) \\ \hline    
M. Adjusted Rand Index                  & $0.91$       & $0.61$       & $0.61$       & $-0.04$       \\                              
M. Adjusted Wallace Index Newcomers 1   & $0.92$       & $0.90$       & $0.45$       & $-0.04$       \\  
M. Adjusted Wallace Index Newcomers 2   & $0.76$       & $0.41$       & $0.72$       & $-0.03$       \\  \hline
\end{tabular}
\end{table}

\textbf{The adjusted original indices}: For each presented type of misclassification of units, the values of all presented indices were calculated (Table~3). As expected, the values of all indices are 1 where the two partitions $U$ and $V$ are the same. Because the Adjusted Rand Index and the Adjusted Fowlkes and Mallows Index are symmetric measures, the values of each of them corresponding to examples (b) and (c) are equal and lower than 1. The values are different in the case of the Adjusted Wallace Index 1 and Adjusted Wallace Index 2. Because the splitting of the two clusters of partition $U$ lowers the value of the Adjusted Wallace Index 1, the value in the case of the example (c) is lower than 1. Similarly, because the merging of the clusters of partition $U$ does not affect the value of the measure, the value in the case of the example (b) is 1. The results are similar for the Adjusted Wallace Index 2: while in the case of the example (b) the value of the index is lower than 1, in the case of the example (c) the value of the index is equal to 1. In the case of two random and independent partitions (example (d)), the values of all indices are close to 0. 

\textbf{The Adjusted Modified Rand Index}: Because the newcomers and outgoers in the case of the Adjusted Modified Rand Index lower the value of the index and consequently indicate the lower stability of the clusters, the value in the example (e) is lower than 1. Since both the splitting and merging of clusters have the same negative effect on the value of the index, the values in both examples (f) and (g) are the same, and lower than in the example (e) which illustrates two partitions with no splitting and merging of clusters. Since in the case of the Adjusted Modified Rand Index (examples (i) to (l)) there are no newcomers which lower the index value, all corresponding values are higher than the values of the Adjusted Modified Rand Index in the examples (e) to (h). The value corresponding to the example (i) is the highest compared with the values of the examples (j) and (k). The latter values are equal due to the already mentioned effect of splitting and merging of clusters on the index value as in the case of the Adjusted Modified Rand Index. Even the values of the Adjusted Modified Rand Index in the examples (m) to (p) are calculated using the same Eq.~(\ref{eq:mri}), the values are the highest compared to the examples (e) to (h) and (i) to (l) due to the fact that, when only outgoers are present, the outgoers lower the value of the index by themselves and also as a consequence of the splitting of the clusters.

\textbf{The modified Wallace indices}: If the splitting and merging of clusters have to be considered differently, according to the operationalization of the stability of clusters, one of the Adjusted Wallace indices can be used. If two partitions are obtained on two different sets of units with a non-empty intersection and if the number of units not present in both sets of units have to be considered as a factor lowering the value of an index, the Adjusted Modified Wallace Index 1 or the Adjusted Modified Wallace Index 2 can be used. When newcomers and outgoers are present, the values of both adjusted modified Wallace indices are lower than 1 even if all other pairs of units are classified in the same clusters in both partitions (example (e)). The merging of clusters does not further lower the value of the Adjusted Modified Wallace Index 1 (example (f)), while the splitting of clusters does (example (g)). On the other hand, the value of the Adjusted Modified Wallace Index 2 equals in both (quasi-identical and splitting) cases since the splitting of clusters does not lower the value of the index.

If no newcomers are present in the second partition, the Adjusted Modified Wallace Index 1 and the Adjusted Modified Wallace Index 2 can be simplified to the Adjusted Modified Wallace Index Outgoers 1 and the Adjusted Modified Wallace Index Outgoers 2. Here, the values are higher than the values of the Adjusted Modified Wallace Index 1 or the Adjusted Modified Wallace Index 2 because there is one factor (newcomers) less that lowers the value of the indices. On the other hand, when there are no outgoers the modified Wallace indices can be simplified as the Adjusted Modified Wallace Index Newcomers 1 or the Adjusted Modified Wallace Index Newcomers 2. Here, the values are higher than the Adjusted Modified Wallace Index 1 and the Adjusted Modified Wallace Index 2 due to the fact that there are two factors (outgoers and the splitting of clusters as a consequence of the outgoers) less that lower the value of the index. Simmilarly, it can be done when there are only newcomers assumed but not outgoers. 

\subsection{The stability of research teams}

The data\footnote{The data are obtained from the Co-operative Online Bibliographic System and Services (COBISS) and the Slovenian Current Research Information System (SICRIS) maintained by the Institute of Information Science (IZUM) and the Slovenian Research Agency (SRA).}, used in this section, are already presented and analyzed by \cite{cugmas2016stability} where the authors aimed to study the stability of the research teams on the level of scientific disciplines in two time periods: 1991-2000 and 2001-2010. The research teams were identified based on the co-authorship networks (in such networks, two researchers are linked if they co-authored at least one scientific bibliographic unit) using generalized blockmodeling \citep{doreian2005}. 

\begin{figure}[H]
    \begin{subfigure}[b]{0.45\textwidth}
        \includegraphics[width=\textwidth]{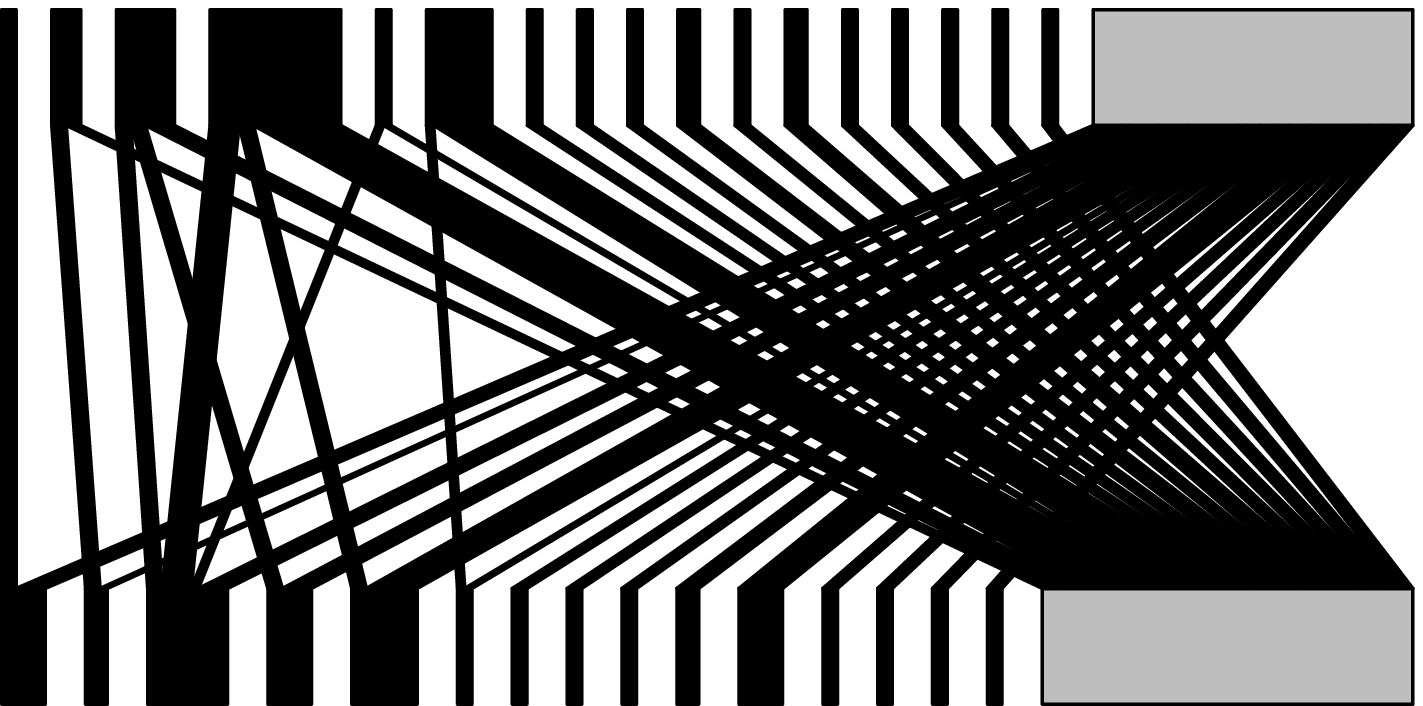}
        \caption{educational studies (N=379)}
        \label{leva}
    \end{subfigure}
    \hfill
    \begin{subfigure}[b]{0.45\textwidth}
        \includegraphics[width=\textwidth]{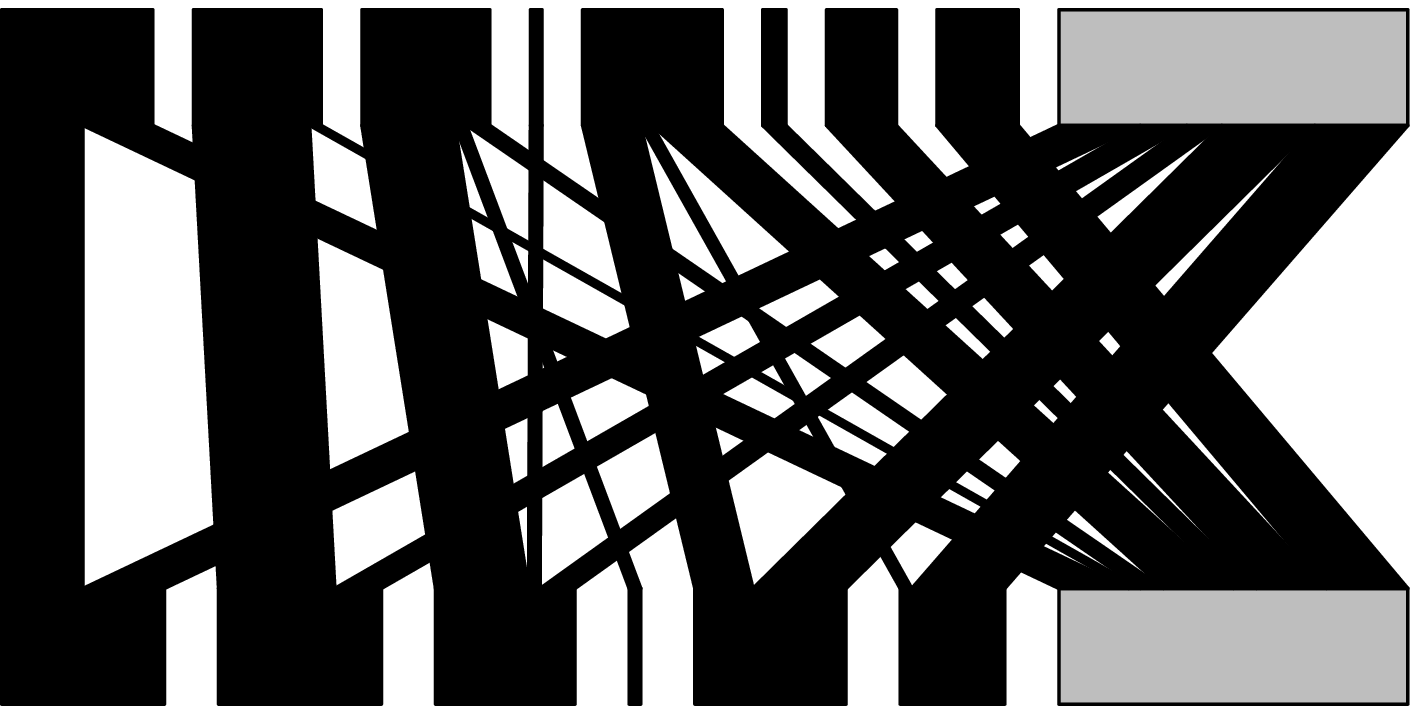}
        \caption{microbiology and immunology (N=226)}
        \label{desna}
    \end{subfigure}
    \caption{Visualizations of researchers' transitions between the research teams (the black rectangles on the top (first time period) and on the bottom (second time period) correspond to the research teams, the gray rectangles on the top correspond to the group of newcomers while the gray rectangles on the bottom correspond to the group of outgoers)}
    \label{examples}
\end{figure}

Although \cite{cugmas2016stability} analyzed most scientific disciplines, only two of them are discussed in this paper. These are "educational studies" and "microbiology and immunology". The clusters of two partitions of two scientific disciplines are visualized by black rectangles on Fig.~\ref{examples}. Some splitting and merging of the research teams can be seen in the case of both scientific disciplines. New researchers can join the already existing research teams (newcomers) or leave them (outgoers). They are denoted by gray rectangles in Fig.~5.

The use of a certain index depends on the data and the operationalization of the stability of the research teams. \cite{cugmas2016stability} operationalized the stability in such a way that the splitting of the research teams and outgoers indicates a lower level of stability. On the other hand, the merging of the research teams and newcomers does not indicate a lower level of research team stability since they indicate a higher level of collaboration among the researchers. Therefore, Modified Wallace Index Outgoers 1 was used in the study of \cite{cugmas2016stability}. 

To illustrate the properties of the proposed indices, the values of all indices are calculated for the partitions of the scientific disciplines "educational studies" and "microbiology and immunology" (Table~4). The newcomers and outgoers are removed from the database when calculating the indices, which do not assume the newcomers or the outgoers. 

When comparing the values of the indices of "educational studies" and "microbiology and immunology", one can observe generally lower values in the case of "educational studies", which can also be noticed by comparing Fig.~5a and Fig.~5b. The values of the original Rand and Wallace indices are generally higher than the values of the modified indices as there are no newcomers and outgoers being considered here. By comparing the Adjusted Wallace Index 1 and Adjusted Wallace Index 2 one can conclude there is a higher level of merging rather than splitting of research teams in "microbiology and immunology", albeit the difference is minor. 

The values of all proposed modified indices are close to 0 in the case of "educational studies" since a very low fraction of researchers is classified in one of the scientific teams in both time periods -- the research teams are quite short termed. However, the values are higher in the case of "microbiology and immunology". Here, the main source of the research teams' instability is the outgoers and newcomers. The comparisons of the Modified Adjusted Wallace Index Outgoers 1 with Modified Adjusted Wallace Index Newcomers 1 and Modified Adjusted Wallace Index Outgoers 2 with Modified Adjusted Wallace Index Newcomers 2 indicates that the newcomers usually join already existing research teams (instead of forming their own teams) in the first period while the outgoers leave the research teams (where decay of the whole research team is unlikely).

\begin{table}[H]
\centering
\caption{The values of different indices for the stability of the research teams for two scientific disciplines}
\label{all_indices}
\begin{tabular}{|l|c|c|}
\hline
                                     & \multicolumn{1}{l|}{educational studies} & \multicolumn{1}{l|}{\begin{tabular}[c]{@{}l@{}}microbiology and \\ immunology\end{tabular}} \\ \hline \hline
M. Adjusted Rand Index               & $0.32$                                     & $0.88$                                                                                        \\ \hline
M. Adjusted Wallace Index 1          & $0.34$                                     & $0.86$                                                                                        \\ \hline
M. Adjusted Wallace Index 2          & $0.31$                                     & $0.91$                                                                                        \\ \hline
M. Adjusted Wallace Index Outgoers 1 & $0.04$                                     & $0.32$                                                                                        \\ \hline
M. Adjusted Wallace Index Outgoers 2 & $0.01$                                     & $0.16$                                                                                        \\ \hline
M. Adjusted Wallace Index Newcomers 1 & $0.01$                                     & $0.16$                                                                                        \\ \hline
M. Adjusted Wallace Index Newcomers 2 & $0.07$                                     & $0.26$                                                                                        \\ \hline
\end{tabular}
\end{table}

\section{Conclusion}

Several well-known indices have been proposed for measuring the similarity of two partitions obtained on one set of units. One of the most often used index is the Rand Index where both the splitting and merging of clusters result in a lower value of the index. Next is the Wallace Index where the merging and splitting of clusters have different effects on the index value.

In comparison to the mentioned indices, the indices newly proposed in this paper assume that two partitions are obtained on two different sets of units where the intersection between the two sets is a non-empty set. Usually, the first set of units is measured at the first time point and the second set at the second time point. All indices can take values on the interval between 0 and 1, where a higher value indicates more stable or similar partitions. The value 1 is only possible when units are present in the first set of units but are not present in the second set of units (outgoers) or units that are not present in the first set of units but are present in the second set of units (newcomers).

The first proposed index is the Modified Rand Index. It is assumed that the number of newcomers and outgoers lowers the value of the index, as does the splitting and merging of clusters. In the case where there are no outgoers or no newcomers, the Modified Rand Index can be simplified.

The next proposed indices are the Modified Wallace Index 1 and the Modified Wallace Index 2. As in the case of the Modified Rand Index, the number of units which are not in the intersection of the two sets of units lowers the value of the index. The splitting of clusters is another factor lowering the index value in the case of the Modified Wallace Index 1 and the merging of clusters in the case of the Modified Wallace Index 2. Depending on the presence of the newcomers and the outgoers, the Modified Wallace Index 1 and the Modified Wallace Index 2 can be simplified as this has been done by \citep{cugmas2016stability} where the Modified Wallace Index Outgoers 1 was used. As described in the introductionary section, the aim is usually to study the stability of research teams that publish scientific bibliographic units together in two time periods. The resutls show that the research teams are relatively unstable in time, which is mainly caused by a high level of outgoers. There are differences in the stability of research teams among the scientific disciplines. 

In the case of the Modified Wallace Index 1, the Modified Wallace Index Outgoers 1  and the Modified Wallace Index 2 (as well as in the case of the Modified Rand Index with outgoers present), the effect of outgoers that lower the value of the index is higher than the effect of the newcomers and splitting of clusters which is not a consequence of the outgoers. Depending on the definition of the stability of clusters, the effect can be corrected. Moreover, some further research could be conducted to define an index where the merging of clusters would increase the index value while the splitting of clusters would decrease the value of an index. Such an index would be useful in some longitudinal studies, e.g. for studying the stability of groups of students in a certain class over the years.

For all presented indices a suggestion for correction for chance is also proposed. The estimation procedure is based on simulations because the correction for chance cannot be obtained on the assumption of fixed frequencies in the contingency table. However, future research could explore the problem of exact correction by chance along with the question of obtaining the confidence intervals analytically.

\end{document}